\documentclass[pdflatex,sn-mathphys-num]{sn-jnl}

\usepackage{graphicx}
\usepackage{multirow}
\usepackage{amsmath,amssymb,amsfonts}
\usepackage{amsthm}
\usepackage{mathrsfs}
\usepackage[title]{appendix}
\usepackage{xcolor}
\usepackage{textcomp}
\usepackage{manyfoot}
\usepackage{booktabs}
\usepackage{algorithm}
\usepackage{algorithmicx}
\usepackage{algpseudocode}
\usepackage{listings}
\usepackage{microtype}

\usepackage{setspace} 
\usepackage[labelfont=bf, labelsep=period]{caption} 
\usepackage{rotating} 
\usepackage{array} 
\usepackage{makecell}  
\usepackage{stackengine} 
\usepackage{xr}
\usepackage{lmodern}

\theoremstyle{thmstyleone}%
\theoremstyle{thmstyletwo}%

\theoremstyle{thmstylethree}%

\begin{document}

\title[Article Title]{A Predictive Design Framework for a Soft Robotic Ventricle using Contractile Actuators}


\author[1]{\fnm{Jeongmin} \sur{Kim}}\email{jk72@illinois.edu}

\author[1]{\fnm{Qiong} \sur{Wang}}\email{qiong.wang@northwestern.edu}

\author[1]{\fnm{Liuyang} \sur{Cheng}}\email{liuyang5@illinois.edu}

\author[1]{\fnm{Samuel} \sur{Tsai}}\email{samuelt7@illinois.edu}

\author[1]{\fnm{Seong Hyeon} \sur{Kim}}\email{shkim71@mit.edu}

\author[3]{\fnm{Harma K.} \sur{Turbendian}}\email{hturbendian@llu.edu}

\author*[1,2]{\fnm{Sameh} \sur{Tawfick}}\email{tawfick@illinois.edu}

\affil[1]{\orgdiv{Department of Mechanical Science and Engineering}, \orgname{University of Illinois Urbana-Champaign}, \orgaddress{\street{1206 W Green St}, \city{Urbana}, \postcode{61801}, \state{Illinois}, \country{USA}}}

\affil[2]{\orgdiv{Beckman Institute for Advanced Science and Technology}, \orgname{University of Illinois Urbana-Champaign}, \orgaddress{\street{405 N Mathews Ave}, \city{Urbana}, \postcode{61801}, \state{Illinois}, \country{USA}}}

\affil[3]{\orgdiv{Department of Cardiothoracic Surgery}, \orgname{Loma Linda University Health}, \orgaddress{\street{11175 Campus Street}, \city{Loma Linda}, \postcode{92354}, \state{California}, \country{USA}}}


\abstract{The natural cardiac cycle is divided into the systole and diastole phases which encompass four distinct stages: isovolumetric contraction and ejection during systole, followed by isovolumetric relaxation and filling during diastole. Cardiovascular modeling of this cycle ranges from high-fidelity multiphysics simulations to reduced-order lumped-parameter (Windkessel) representation of the heart--artery coupling. However, current models do not relate the mechanics of the actuator driving the ventricle pump to the hemodynamics. In this study, we develop and experimentally validate a predictive design framework for ventricle-like pumps using various types of soft contractile actuators. We build a circulatory loop which reproduces the entire loop including the isovolumetric phases—where pressure changes occur without volume shifts. The framework is based on a lumped-parameter model, hereafter referred to as the phase-dependent Actuator-driven Windkessel 3-element (AWK3) model, to bridge soft actuator mechanics to the cardiac pressure-volume (P--V) loop. Unlike traditional models that require either pressure or volume as a fixed input to estimate the other, our proposed model predicts both variables when informed by the isometric characteristics of the actuators. We validate the model using a ventricle-inspired pump driven by  a linear contractile series-elastic actuator or twisted and coiled polymer actuators (TCPA). We relate the actuator isometric testing protocol to the phase-dependent AWK3 model, which  replicates the Frank-Starling law, accurately describing cardiac behavior under varying conditions of preload, afterload, and inotropy (contractility). This approach provides a robust platform for the design and high-fidelity control of bio-inspired soft robotic circulatory systems.}




\maketitle

\section{Introduction}\label{Section1}

Congenital heart disease (CHD) and progressive heart failure often require mechanical circulatory support systems, such as ventricular assistive devices (VADs), as a critical bridge to transplantation or long-term therapy \cite{van2011birth, hoffman2002incidence, wren2012epidemiology, marelli2014lifetime, agopian2017genome, foroughi2026soft}. Currently, only a limited number of FDA-approved artificial hearts exist, with devices like the Berlin Heart EXCOR remaining the standard for infants and small children \cite{almond2013berlin, copeland2012experience}. However, these traditional devices rely on rigid pneumatic pulsatile or continuous axial flow mechanisms that necessitate lifelong systemic anticoagulation, introducing severe clinical complications \cite{copeland2012experience, ott2025impact}. To overcome the limitations of these rigid VADs and rigorously evaluate novel cardiovascular implants, there is a growing demand for high-fidelity \textit{in vitro} testing platforms \cite{ji2026soft}. Specifically, heart-inspired soft bioreactors and benchtop simulators are required to accurately mimic the anisotropic compliance and hemodynamics of the native myocardium \cite{weymann2023artificial, rosalia2023pneumatic, park2024biorobotic, davies2024soft}. However, the myocardium is a multifunctional tissue with locally tailored anisotropic biomechanics, which enables the various functions during each of the cardiac cycle stages.   

A crucial requirement for mimicking the heart muscle function is replicating the active-passive mechanical coupling between contractile elements and structural tissues, respectively, even in a single direction \cite{granzier2004giant, linke2008sense}. While recent biohybrid hearts have successfully integrated synthetic actuators with biological tissues to serve as high-fidelity simulators capturing these complex mechanics \cite{park2024biorobotic}, engineering this delicate interplay within fully synthetic soft robotic systems remains a challenge. This coupling allows the system to emulate key physiological behaviors, including the Frank-Starling mechanism where increased stroke volume results from higher preload, the dynamic adaptation to varying afterloads, and modulation of the intrinsic contractility.  The interactions between the active and passive behavior result from the cyclic requirements which not only include the geometrically tailored contractile motions during ejection but also the highly anisotropic passive tissue expansion during filling. Since anisotropically tailored soft engineering material cannot replicate the myocardium, current approaches in soft robotic cardiovascular systems predominantly rely on simple uniaxial actuators coupled to isotropic elastomeric materials such as using soft pneumatic actuators \cite{roche2017soft, guillen20263d}. While these systems have demonstrated remarkable physiological adaptability, their design approach still relies on non-systematic trial-and-error methodologies for tuning these mechanics \cite{zrinscak2025design}. There is a need for a framework for systematic selection and tailoring of the passive and active responses of soft actuators based on reliable modeling of the heart pumping cycle \cite{park2022computational}.

To bridge the gap between tissue mimicry and mechanical circulatory support, recent efforts have increasingly focused on developing soft robotic benchtop soft pump simulators. Notable advancements include pneumatically actuated patient-specific hydrodynamic models \cite{rosalia2023pneumatic}, biorobotic hybrid hearts incorporating soft components to serve as mitral valve simulators \cite{park2024biorobotic}, and hydraulically driven soft ventricular analogs capable of reproducing complex myocardial biomechanics such as apical torsion \cite{davies2024soft}. While these platforms successfully generate physiological pulsatile flow waveforms and mimic macroscopic cardiac motions, their functional fidelity is heavily constrained by the mechanical limits of their underlying artificial muscles. To drive such bio-inspired systems, twisted-and-coiled polymer actuators (TCPAs) have emerged as promising candidates due to their contractile motion, exceptional power density, scalability, and inherent compliance \cite{haines2014artificial, hu2024artificial, witham2024coil, almubarak2017twisted}. However, despite extensive studies, it remains unclear whether TCPAs have the stroke and force capabilities required for heart pumping. The vast majority of current characterizations focus predominantly on isobaric metrics—measuring stroke displacement under constant loads—or assessing cyclic endurance \cite{wang2023insect, ren2022stepwise, tsai2025high, chen2024effect, zhang2024compound}. This conventional benchmarking cannot capture the precise mechanics of the isovolumetric phases, where extreme pressure shifts must occur rapidly without volume changes. Without characterizing these constant-strain behaviors, existing simulators cannot select suitable actuators based on a predictive design framework. Specifically, a soft actuator selection framework for heart functions should fully map the dynamic active-passive mechanical coupling required to accurately replicate the Frank-Starling law, thereby leaving a crucial gap in comprehensive physiological emulation.

\begin{figure}[tp!]
    \centering
    \includegraphics[width=1\linewidth]{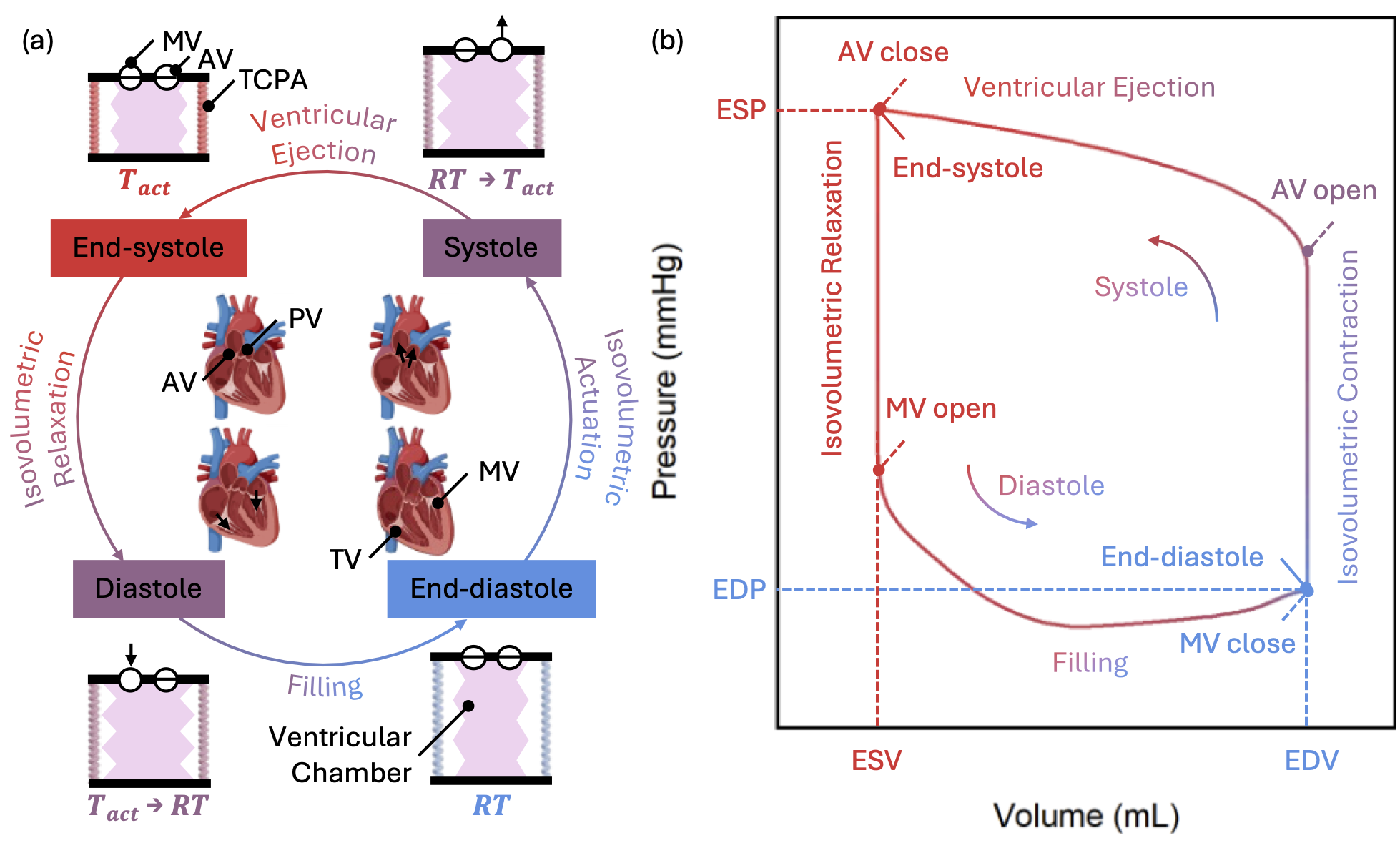}
    \caption{\textbf{Operational principle of the twisted and coiled polymer actuator (TCPA) driven soft robotic ventricle.} (a) Schematic of cardiac cycles in the natural hearts and soft robotic pump. The four phases of the cycle—filling, isovolumetric contraction (IVC), ventricular ejection, and isovolumetric relaxation (IVR)—are regulated by closing and opening of the valves: inlet valve same as mitral valve (MV) at the left ventricle and tricuspid valve (TV) at the right ventricle, and outlet valve same as aortic valve (AV) at the left ventricle and pulmonary valve (PV) at the right ventricle. Systole is achieved by thermal actuation by increasing temperature of TCPAs from room temperature ($RT$) to actuation temperature ($T_\text{act}$). The temperature of TCPAs reaches $T_\text{act}$ at end-systole (ES) and $RT$ at end-diastole (ED). (b) Pressure-Volume profile of a left ventricle across four cardiac phases. It reaches end-systole pressure (ESP) and volume (ESV) at ES, while returning to end-diastolic pressure (EDP) and volume (EDV) at ED.}
    \label{fig:Figure1}
\end{figure}

In the biological heart, the cardiac cycle is governed by the coordinated opening and closing of four primary cardiac valves: the mitral valve (MV), aortic valve (AV), tricuspid valve (TV), and pulmonary valve (PV) (Fig.~\ref{fig:Figure1}a). The cycle is initiated by the isovolumetric contraction (IVC) phase, where the closure of the MV allows the ventricular myocardium to generate rapid tension while the ventricular volume remains constant. This significant rise in intraventricular pressure marks the transition from the end-diastolic volume (EDV) and end-diastolic pressure (EDP) state toward the onset of ventricular ejection. Subsequently, the AV opens, and the heart ejects blood until it reaches the end-systolic volume (ESV) and end-systolic pressure (ESP) limits. Following ejection, the heart undergoes isovolumetric relaxation (IVR), a phase where all valves remain closed and ventricular pressure drops rapidly without any change in volume. During the subsequent diastolic filling phase, ventricular volume increases significantly while pressure remains relatively constant, returning the heart to its initial end-diastolic state. This sequence of events defines the macroscopic pressure-volume (P-V) profile shown in Fig.~\ref{fig:Figure1}b. For artificial hearts driven by contractile actuators, replicating this isovolumetric sequence is paramount, as it serves as the critical transition bridging intrinsic actuator mechanics with these macroscopic P-V profiles. 
The fundamental root of these challenges lies in the "missing link" of the cardiac cycle: the IVC and relaxation phases. During these critical periods, the actuators must generate rapid shifts in internal force while maintaining a constant fluid volume. While the thermomechanical physics of TCPAs are becoming increasingly well-understood at the individual muscle level \cite{wang2024mechanics}, mathematically translating this mechanical behavior into continuous system-level hemodynamics remains an unresolved challenge. On one hand, multiphysics cardiovascular modeling with fluid-structure interactions and hemodynamics enable basic disease research, \textit{in silico} drug evaluation, and precision medicine \cite{kim2010patient, taylor2009patient, nordsletten2011coupling, nash2000computational, bazilevs2009computational}. On the other hand, to efficiently describe the interplay between fluid transport and mechanical load, reduced-order lumped parameter models, such as the Windkessel (WK) model, are used to model the heart pump and its interactions with arteries \cite{segers2003systemic, kim2009coupling}. These models effectively characterize the relationship between systemic arterial resistance and compliance in a circulatory loop, allowing researchers to estimate global hemodynamics \cite{sagawa1990translation, westerhof2009arterial}. However, existing modeling frameworks lack a direct and predictive connection between the active mechanical state of the contractile actuator and the resulting P-V output \cite{arfaee2025soft, rocchi2024patient, demeersseman2025activation}. Specifically, WK architectures typically require either pressure or volume as a pre-defined, fixed input boundary condition to estimate the complementary variable. Consequently, they completely fail to account for how the physical limits of the actuator’s inherent force generation—particularly during the highly demanding isovolumetric phases—simultaneously and autonomously govern both pressure and volume within a mechanically coupled fluidic environment.

In this study, we address these critical gaps by proposing a novel 3-element reduced-order framework, which we will call the phase-dependent Actuator-driven Windkessel (AWK3) model in this study, to translate actuator mechanics into soft robotic ventricle performance. Our core idea is to establish a systematic mathematical relationship between the characteristics of the actuators and the resulting Pressure-Volume (P-V) cycle. It should be noted that this model is phase-dependent, as the systolic and diastolic phases are governed by distinct physical mechanisms; systole is actively driven by the actuation unit, whereas diastole undergoes passive recovery. To construct this relationship, we first establish and rigorously validate the foundational model using a Series Elastic Actuator (SEA). By utilizing the SEA, we can explicitly parameterize the active-passive mechanical coupling and precise electromechanical limitations within a benchtop simulator. Building upon this highly controlled baseline, we subsequently apply the phase-dependent AWK3 framework to soft artificial muscles using the isometric test to characterize the passive, active, and total force components of TCPAs across an extensive strain range. By integrating these muscle-level isometric profiles into the established model, we consider the isovolumetric phase within the design framework. This approach enables the simultaneous prediction of both pressure and volume without requiring either as a fixed boundary condition. By replicating the Frank-Starling mechanism under varying preload and afterload conditions across both SEA and TCPA systems, this research provides a robust, systematic methodology to replace trial-and-error approaches in the design, selection, and evaluation of contractile actuators for next-generation cardiac bioreactors.

\section{Materials and Methods}

\subsection{Phase-dependent AWK3 model Design Framework}
We developed the phase-dependent AWK3 model and use it to construct a reduced-order design framework for soft heart-inspired pumps. The WK model predicts various stages of the heart cycles with a simple system of differential equations using lumped dissipative and storage elements such as  resistors and capacitors, respectively \cite{rosalia2024modulating}. The design framework can be extended to more detailed computational tools which capture more physiological elements via finite element methods, computational fluid dynamics \cite{marsden2014optimization}, or coupled fluid structure interactions \cite{bucelli2023mathematical, bonini2026monolithic}. We couple the passive and active actuator mechanics to the WK model to mathematically bridge the empirical isometric force limits of soft actuators to the macroscopic fluidic boundaries of the circulatory loop. By establishing this active-passive mechanical coupling, the required inotropy, preload, and afterload conditions can be systematically predicted and tuned prior to physical system integration. 

Based on this strategy, we structured our study into four phases: (1) mathematical formulation of the phase-dependent AWK3 model, (2) parameter identification and baseline validation using an SEA-driven mock circulatory loop (MCL), (3) isometric characterization of the isolated TCPAs, and (4) closed-loop dynamic evaluation of the fully integrated soft robotic pump.

\subsection{Mock Circulatory Loop}
\begin{figure}[tp!]
    \centering
    \includegraphics[width=1\linewidth]{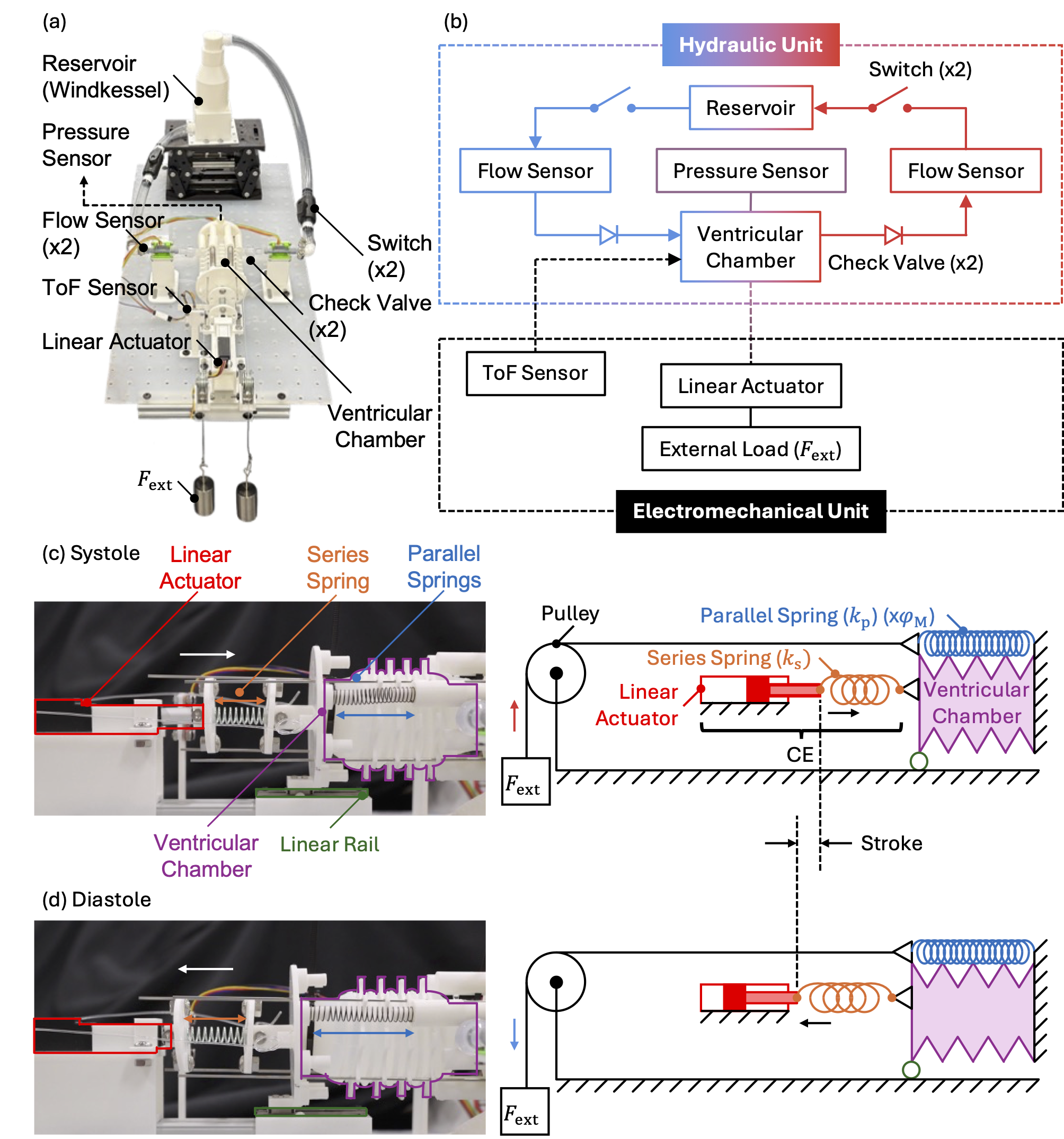}
    \caption{\textbf{Design and modeling of the Mock Circulatory Loop (MCL) for design framework validation.} (a) Experimental setup of the MCL incorporating the soft robotic ventricle, data acquisition system including a Time-of-Flight (ToF) sensor, and external load ($F_\text{ext}$). (b) System architecture illustrating the interaction between the hydraulic unit and electromechanical unit. (c--d) Photograph and schematic representations of the systole and diastole modes, highlighting the coordination between the linear actuator, series and parallel compression springs (with stiffness $k_s$ and $k_p$, respectively), and ventricular chamber.}
    \label{fig:Figure2}
\end{figure}

To identify the values of the lumped hydrodynamic model parameters of the soft pump, we constructed a mock circulatory loop (MCL) comprising a hydraulic unit and a mechanical unit (Fig.~\ref{fig:Figure2}a--b). The hydraulic unit emulates the systemic circulation through four key components: (i) a bellows-shaped ventricular chamber fabricated from silicone elastomer (Ecoflex 00-30, Smooth-On Inc., USA) to represent the heart's structure, (ii) a rigid reservoir acting as a WK element to dampen pressure fluctuations, (iii) one-way check valves (91050, Qosina, USA) to ensure unidirectional flow, and (iv) a vascular tubing network. To maintain physiological viscosity, we utilized a 40\,wt\% glycerin and 60\,wt\% water mixture.

The mechanical unit integrates the actuators with the ventricular chamber to drive cyclic flow. As illustrated in Fig.~\ref{fig:Figure2}c–d, unidirectional flow is regulated by the bidirectional motion of the actuators throughout the cardiac cycle. During systole, the contractile unit compresses the ventricular chamber to eject fluid into the systemic outlet circuit. In the subsequent diastole, the actuator relaxes or retracts, allowing the ventricular chamber to expand and refill via hydrostatic pressure from the reservoir. To monitor these dynamics in real-time, a board-mounted pressure sensor (SSCDANV030PAAA5, Honeywell, USA) and in-line liquid flow sensors (SLF3S-4000B, Sensirion, Switzerland) were utilized to track hemodynamic performance, while a Time of Flight (ToF) displacement sensor (VL53L4CD, SparkFun Electronics, USA) measured the displacement of the ventricular chamber.

\subsection{SEA-Driven Testing of the Soft Robotic Ventricle}
First, we used an electromechanical linear actuator with a spring in series, known as series elastic actuator (SEA), to identify the parameters of the phase-dependent AWK3 model using the MCL. In this setup, the active mechanical unit of the MCL consisted of the linear actuator and a series spring ($k_\text{s} = 2452.0\,\text{N/m}$) to mimic the behavior of a soft actuator. This linear actuator provided precisely controlled bidirectional motion. Furthermore, parallel springs ($k_\text{p} = 141.6\,\text{N/m}$) were incorporated to intentionally mimic the passive stiffness inherent to the TCPAs. This electromechanical platform allowed us to explicitly parameterize the active-passive mechanical coupling and evaluate the core mathematical relationships of the phase-dependent AWK3 model without the nonlinear complexities associated with TCPA actuation. To establish this baseline, we first operated the SEA-driven pump under a standard cardiac cycle to validate the fundamental mathematical framework of the phase-dependent AWK3 model (Fig.~\ref{fig:Figure3}). By continuously measuring the stroke, pressure, and volume trajectories across the four cardiac phases, we compared the predictive capabilities of our phase-dependent AWK3 model against the conventional WK3 model. 

Within this framework, we systematically replicated the Frank-Starling mechanism by modulating three key physiological determinants:
(i) Inotropy (Contractility): controlled by the duty cycle of the linear actuator, the number of integrated parallel springs ($\varphi_\text{M}$), and the magnitude of external loads ($F_\text{ext}$); 
(ii) Preload: regulated by adjusting the initial volume of the ventricular chamber; and 
(iii) Afterload: modulated by varying the degree of outlet valve constriction to simulate systemic vascular resistance. 
This controlled environment provided a robust baseline for evaluating the simultaneous pressure-volume co-prediction capability of the phase-dependent AWK3 model.

\subsection{Isometric Characterization of TCPAs}
Following the baseline validation, we transitioned to soft artificial muscles to replicate bio-inspired pumping mechanics. Detailed fabrication procedures are provided in Appendix~\ref{Appendix1}. The final geometries of the fabricated TCPAs are summarized in Table~\ref{Table1}.

\begin{table}[t]
    \centering
    \footnotesize
    \renewcommand{\arraystretch}{1.2} 
    \caption{\textbf{List of TCPAs and their properties.} The specifications include fiber diameter, structure type, mandrel size, spring index, heating method, heating wire diameter, and actuation temperature.}
    \label{Table1}
    \begin{tabular*}{\textwidth}{@{\extracolsep{\fill}} l c c c c c c c }
        \toprule
        \textbf{Name} & \textbf{Fiber} & \textbf{Structure} & \textbf{Mandrel} & \textbf{Spring} & \textbf{Heating} & \textbf{Heating} & \textbf{Temp} \\
                      & \textbf{(mm)}  &                    & \textbf{(mm)}    & \textbf{Index}  & \textbf{Method}  & \textbf{Wire (mm)} & \textbf{($^\circ$C)} \\
        \midrule
        M-1-0.75    & 1.0  & Monocoil  & 0.75 & 1.75 & Heat gun      & -     & 100 \\
        M-1-1.15     & 1.0  & Monocoil  & 1.15 & 2.15 & Heat gun      & -     & 100 \\
        M-1-2.33    & 1.0  & Monocoil  & 2.33 & 3.33 & Heat gun      & -     & 100 \\
        M-1.3-0.75  & 1.3  & Monocoil  & 0.75 & 1.58 & Heat gun      & -     & 100 \\
        M-1.3-1.15   & 1.3  & Monocoil  & 1.15 & 1.88 & Heat gun      & -     & 100 \\
        M-1.3-2.33  & 1.3  & Monocoil  & 2.33 & 2.79 & Heat gun      & -     & 100 \\
        S-0.47-0.75 & 0.47 & Supercoil & 0.75 & 1.63 & Joule heating & 0.1   & 150 \\
        S-0.47-1.5  & 0.47 & Supercoil & 1.50 & 2.26 & Joule heating & 0.1   & 150 \\
        S-0.15-0.25 & 0.15 & Supercoil & 0.25 & 1.37 & Joule heating & 0.075 & 100 \\
        H-0.47-0.32 & 0.47 & Hypercoil & 0.32 & 1.16 & Joule heating & 0.1   & 160 \\
        \bottomrule
    \end{tabular*}
\end{table}

To accurately parameterize the phase-dependent AWK3 model which must account for both isovolumetric and isobaric cardiac phases, we established an isometric testing framework to characterize mechanical outputs, such as force generation and contractile stroke. The proximal end of the actuator was secured to a force/torque sensor (SI-8-0.05, ATI Industrial Automation, USA), while the distal end was fixed to an adjustable translation stage. Zero-strain equilibrium was defined when the measured force balanced with the weight of the actuator. Before data acquisition, thermal training was performed via 20 heating-cooling cycles, with the maximum temperature limited to 80\% of the intended testing maximum \cite{peng2018study, aziz2023plant}.

Tests were conducted at 0.5 mm strain increments. Monocoiled TCPAs were actuated symmetrically using opposed heat guns, whereas supercoiled and hypercoiled TCPAs were actuated via Joule heating through the integrated electrical wires.

\subsection{TCPA-Driven Testing of the Soft Robotic Ventricle}

Next, we replaced the rigid SEA system with TCPAs. Although they possess different force generation capacities and actuation speeds, we accommodated the inherent passive stiffness of the TCPAs by eliminating the external parallel springs from the MCL. Furthermore, to tailor this active-passive coupling, the hydraulic conduits of the MCL were geometrically shortened (Fig.~\ref{fig:FigureS6}a). This adjustment effectively reduced the systemic outlet resistance, ensuring that the fluidic impedance was perfectly matched to the contractile limits of the soft artificial muscles. In this configuration, the TCPAs were directly mounted in parallel with the ventricular chamber. The electrical wires of the 4 TCPAs were connected in series with each other using Wago connectors to ensure uniform electrical distribution. To drive the contraction, the TCPAs were Joule-heated with a constant current of 0.45 A, reaching a target actuation temperature of 150 $^\circ$C regardless of the number of integrated modules. The cardiac cycle duration was set to 30 s with a 30\% systolic duty cycle, maintaining consistency with the SEA-driven baseline. Furthermore, symmetrically positioned fans were employed at the sides of the system to accelerate the convective cooling of the TCPAs during the diastolic phase.

\section{Results}

\subsection{Governing Equations of the Soft Robotic Ventricle}

We develop a quantitative design framework for a soft robotic heart based on a lumped-parameter modeling of the natural cardiac P-V loop, which consists of diastolic filling, IVC, ventricular ejection, and IVR phases (Fig.~\ref{fig:Figure1}). The hydraulic network alternates between systole and diastole over the total duration of $t_\text{cycle}$.

\begin{figure}[tp!]
    \centering
    \includegraphics[width=0.5\textwidth]{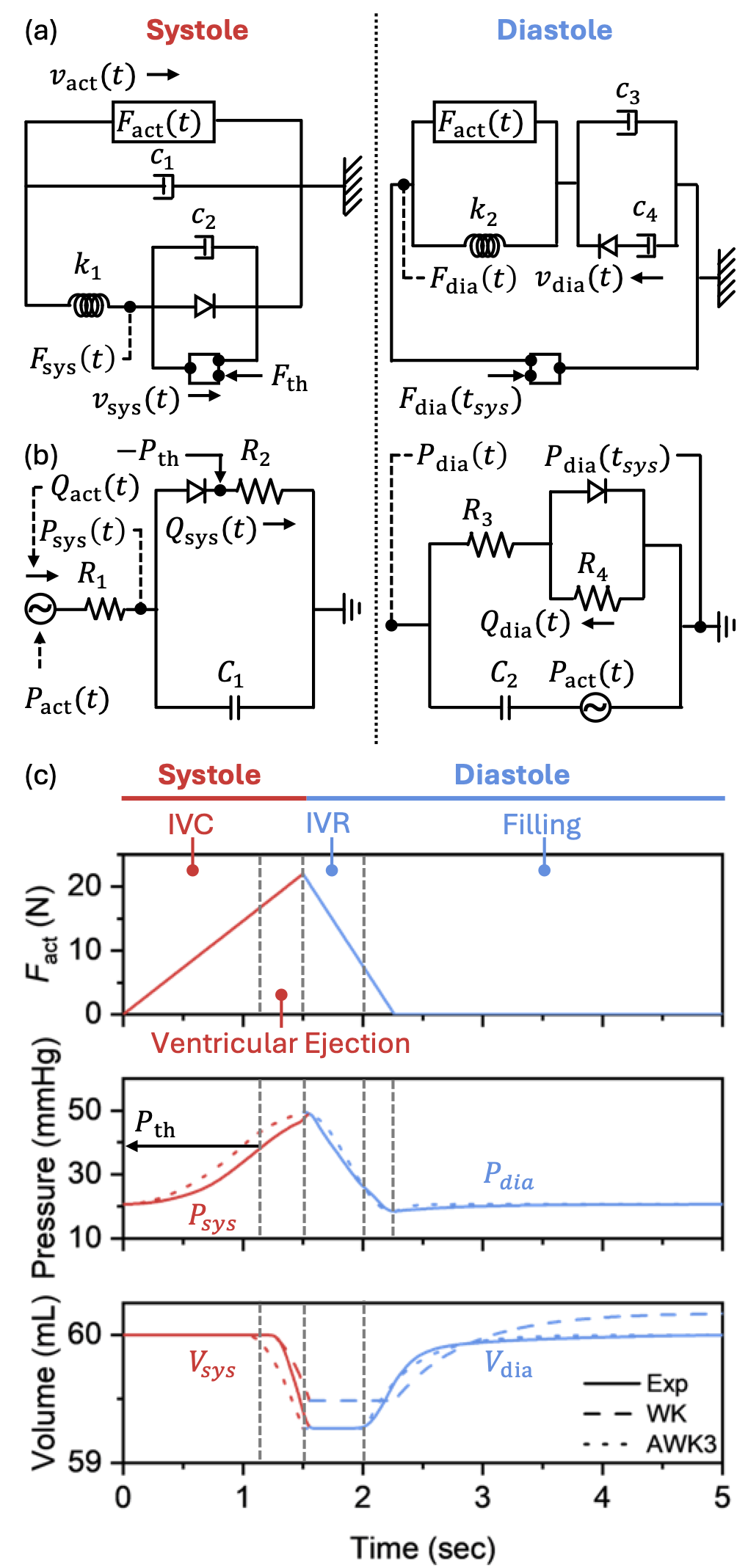}
    \caption{\textbf{Phase-dependent actuator-driven Windkessel 3-element model (AWK3) and experimental validation.} (a) Mechanical free-body diagrams of the systolic and diastolic phases. (b) Equivalent hydraulic circuits derived from the mechanical models. (c) Model input-output comparison: pressure input for the Windkessel (WK) model versus actuator force input for the phase-dependent AWK3 model, alongside the resulting experimental (Exp) measured and predicted volume trajectories. IVC is defined where systolic pressure ($P_\text{sys}$) is lower than threshold pressure ($P_\text{th}$). On the other hand, IVR is defined within the time range of $t_\text{IVR}$ where diastolic pressure ($P_\text{dia}$) is still higher than the initial pressure ($P_\text{sys}(t_\text{cycle})$).}
    \label{fig:Figure3}
\end{figure}

During systole ($0 \le t < t_\text{sys}$), the heart is modeled using a 3-element WK model (Fig.~\ref{fig:Figure3}a--b). We first derived the governing equation directly from the mechanical free-body diagram to verify physical consistency (Fig.~\ref{fig:Figure3}a). This actuation unit, representing the heart muscle ($F_\text{act}(t)$ [N]), is connected in series with the WK3 model, which consists of a dashpot ($c_\text{1}$ [N$\cdot$s/m]) in parallel with a Maxwell element comprising a dashpot $c_\text{2}$ [N$\cdot$s/m] and a spring ($k_\text{1}$ [N/m]). The ventricular force, $F_\text{sys}$, is measured after the spring within the Maxwell branch and represents the driving force increase at the exit of the ventricular chamber. The corresponding ventricular velocity is measured across $c_\text{2}$, as this composition mechanically represents the chamber itself, while $c_\text{2}$ captures the effect of the systemic resistance. Flow is restricted to a single direction by a one-way check valve and is governed by the difference between $F_\text{sys}(t)$ and a threshold force, $F_\text{th}$. $F_\text{th}$ accounts for the combined resistance opposing blood flow, such as the cracking pressure of the valve, and the hydrostatic pressure of the reservoir. By applying Kirchhoff’s analogy, force balance and compatibility equations are defined as:

\begin{equation}
F_{\text{act}}(t) = c_1 v_{\text{act}}(t) + F_{\text{sys}}(t)
\label{Equation_01}
\end{equation}

\begin{equation}
v_\text{sys}(t) = 
\begin{cases} 
0 & \text{if } F_\text{sys}(t) < F_\text{th}\\
\frac{F_\text{sys}(t) - F_\text{th}}{c_\text{2}} & \text{if } F_\text{sys}(t) \ge F_\text{th}
\end{cases}
\label{Equation_02}
\end{equation}

Eq.~\ref{Equation_02} can be simplified as the following:

\begin{equation}
v_{\text{sys}}(t) = \max\left(0, \frac{F_{\text{sys}}(t) - F_{\text{th}}}{c_2}\right)
\label{Equation_03}
\end{equation}

By combining and differentiating Eq.~\ref{Equation_03} and Eq.~\ref{Equation_01}, we obtain the mechanical governing equation:

\begin{equation}
F_{\text{act}}(t) = F_{\text{sys}}(t) + \frac{c_1}{k_1}\frac{dF_{\text{sys}}(t)}{dt} + c_1 \max\left(0, \frac{F_{\text{sys}}(t) - F_{\text{th}}}{c_2}\right)
\label{Equation_04}
\end{equation}

Translating this mechanical system into the analogous  hydraulic circuit, the force and velocity are replaced with pressure $P$ and flow rate $Q$, respectively. The driving force $F_\text{act}$ acts as the pressure source ($P_\text{act}$ [mmHg]). Mechanically, $c_\text{1}$ translates to the systolic proximal resistance $R_\text{1}$ [$\text{mmHg}\cdot\text{s/mL}$], while the Maxwell element ($c_\text{2}-k_\text{1}$) represents the distal resistance and systolic compliance ($R_\text{2}$ [$\text{mmHg}\cdot\text{s/mL}$], $C_\text{1}$ [mL/mmHg]). Note that in this mechanical to hydraulic circuit analogy, the parallel and series connections are inverted. The inlet one-way valve (representing the mitral valve) remains closed while the outlet one-way valve (representing the aortic valve) opens when the ventricular pressure ($P_\text{sys}$ [mmHg]) exceeds the threshold pressure ($P_\text{th}$ [mmHg]). This results in the flow rate ($Q_\text{sys}$ [mL/s]) during the two phases of systole: IVC and ventricular ejection. The corresponding systolic flow rate $Q_\text{sys}(t)$ is defined as:

\begin{equation}
Q_\text{sys}(t) = 
\begin{cases} 
0 & \text{if } P_\text{sys}(t) < P_\text{th} \quad \text{(IVC)} \\
\frac{P_\text{sys}(t) - P_\text{th}}{R_\text{2}} & \text{if } P_\text{sys}(t) \ge P_\text{th} \quad \text{(Ventricular Ejection)}
\end{cases}
\label{Equation_05}
\end{equation}

While the piecewise definition in Eq. \ref{Equation_05} clearly distinguishes the physical phases of IVC and ventricular ejection, it can be equivalently expressed as:

\begin{equation}
Q_\text{sys}(t) = \text{max}\left(0, \frac{P_\text{sys}(t) - P_\text{th}}{R_\text{2}}\right)
\label{Equation_06}
\end{equation}

By applying this transformation to the mechanical free-body diagram (Eq.~\ref{Equation_04}) and incorporating Eq.~\ref{Equation_06}, the lumped-parameter framework yields the pressure governing equation:

\begin{equation}
P_{\text{sys}}(t) = P_{\text{act}}(t) - R_1 C_1 \frac{dP_{\text{sys}}(t)}{dt} - R_1 \max\left(0, \frac{P_{\text{sys}}(t) - P_{\text{th}}}{R_2}\right)
\label{Equation_07}
\end{equation}

The variables $Q_\text{sys}$ and $P_\text{sys}$ can be experimentally measured using the sensors (SSCDANV030PAAA5 and SLF3S-4000B) in the MCL. The threshold pressure, $P_\text{th}$, was determined through systematic calibration and extracted from the pressure at which $Q_\text{sys}$ becomes a non-zero value. The actuation pressure, $P_\text{act}$, was derived based on the contractile element, which is described in detail in Sections~\ref{Section3-2} and \ref{Section3-4}. External load $F_\text{ext}$ will be included in $P_\text{act}$ as it is a resisting force to the actuation unit. An example of $F_\text{act}$ is shown in Fig.~\ref{fig:Figure3}c. By implementing Eqs.~\ref{Equation_06}--\ref{Equation_07}, a simultaneous optimization of the systolic parameters, including $R_\text{1}$, $R_\text{2}$, and $C_\text{1}$, was conducted to minimize the error between the simulated $Q_\text{sys}$ and $P_\text{sys}$ and their experimentally measured values. The optimized parameters are listed in Table \ref{Table2}. The traditional WK3 model allows the calculation of the volumetric output during the systole (IVC and ventricular ejection) (Fig.~\ref{fig:Figure3}c) using the equation:

\begin{equation}
    Q_{\text{sys}}(t) \left(1 + \frac{R_1}{R_2}\right) + C_1 R_1 \frac{dQ_{\text{sys}}(t)}{dt} = \frac{P_{\text{sys}}(t)}{R_2} + C_1 \frac{dP_{\text{sys}}(t)}{dt}
    \label{Equation_08}
\end{equation}

However, the traditional WK3 model relies exclusively on using experimental pressure data $P_\text{sys}$ to calculate the flow rate $Q_\text{sys}$. Mathematically, solving Eq.~\ref{Equation_08} requires calculating the time derivative of this experimental signal ($dP_\text{sys}/dt$), which inherently amplifies measurement noise. When $Q_\text{sys}$ is subsequently integrated over time to estimate the chamber volume, these numerical errors accumulate. Furthermore, the standard WK3 circuit lacks physical boundary conditions, such as the mechanical end-stop of the ventricular chamber and the precise flow restriction by the valve threshold ($P_\text{th}$). These combined mathematical and physical limitations explain why the WK3 model deviates from the experimental baseline, resulting in a misaligned volume overshoot (Fig.~\ref{fig:Figure3}c). In contrast, the phase-dependent AWK3 model developed in this study uses the computed actuation force (top panel) to calculate both the pressure $P_\text{sys}$ and flow rate $Q_\text{sys}$ as detailed in Section~\ref{Section3-2} and \ref{Section3-4}. By directly coupling the state-dependent actuator model with the fluidic network, the phase-dependent AWK3 model predicts the resulting pressure and volume dynamics using the actuator design parameters ($P_\text{act}$) and the lumped fluidic circuit resistance and capacitance (Eq.~\ref{Equation_05}--\ref{Equation_07}). Notably, the volumetric trajectories in the bottom panel demonstrate that while the WK3 model deviates from the experimental baseline (Exp)—underestimating the stroke volume during ejection and causing time delays during the diastolic filling phase. On the other hand, the phase-dependent AWK3 model accurately replicates the experimental volume curve throughout the entire cycle.

During diastole ($t_\text{sys} \le t < t_\text{cycle}$), the force dynamics are governed by the rapid relaxation of the actuator in the early stage, whereas the velocity depends on the instantaneous force, thereby yielding the isovolumetric relaxation (IVR) and filling phases. To accurately capture the distinct behaviors of force and velocity recovery, the governing equations transition dynamically based on the state of the active relaxation force $F_\text{act}(t)$ and the instantaneous diastolic force relative to the end-diastolic reference force $F_\text{dia}(t_\text{cycle})$ (Fig.~\ref{fig:Figure3}a). 

The macroscopic diastolic velocity $v_\text{dia}(t)$ is evaluated across $c_4$, while the diastolic force $F_\text{dia}(t)$ is measured locally at the node between $c_3||k_2$ and $c_4$. The additional damping $c_4$ becomes active when $F_\text{dia}(t)$ drops below the reference force $F_\text{dia}(t_\text{cycle})$, thereby mediating the incoming flow. Therefore, $v_\text{dia}(t)$ is formulated as:

\begin{equation}
v_\text{dia}(t) =
\begin{cases}
    0 & \text{if } F_\text{dia}(t) \ge F_\text{dia}(t_\text{cycle}) \\
    \frac{F_\text{dia}(t_\text{cycle}) - F_\text{dia}(t)}{c_4} & \text{if } F_\text{dia}(t) < F_\text{dia}(t_\text{cycle}) 
\end{cases}
\label{Equation_09}
\end{equation}

This mathematical framework intrinsically preserves the decoupled dynamics of the system. For instance, during the IVR phase, the volume remains constant, dictating that the macroscopic ventricular filling velocity is identically zero. The active relaxation is modeled by directly coupling the rate of force change to the temporal decay of the actuator ($\frac{dF_\text{act}(t)}{dt}$). Applying these physical boundary conditions yields the comprehensive governing equation for the rate of force change:

\begin{equation}
\frac{dF_\text{dia}(t)}{dt} = \frac{dF_\text{act}(t)}{dt} + k_2 \left[ v_\text{dia}(t) - \frac{F_\text{dia}(t)}{c_3} \right]
\label{Equation_10}
\end{equation}

This mechanical relationship is directly converted into the equivalent hydraulic circuit (Fig.~\ref{fig:Figure3}b) by mapping the mechanical parallel connections to hydraulic series connections, substituting velocity with flow rate. Specifically, the variables $F_\text{dia}$ and $v_\text{dia}$ translate into pressure ($P_\text{dia}$) and flow rate ($Q_\text{dia}$), while the parameters $c_3$, $k_2$, and $c_4$ correspond to the refilling resistance $R_3$, diastolic elastance (the inverse of diastolic compliance, $1/C_2$), and viscous damping resistance $R_4$, respectively. 

The diastolic refilling flow $Q_\text{dia}(t)$ is defined as follows:

\begin{equation}
Q_\text{dia}(t) = 
\begin{cases} 
    0 & \text{if } P_\text{dia}(t) \ge P_\text{dia}(t_\text{cycle}) \quad \text{(IVR)} \\
    \frac{P_\text{dia}(t_\text{cycle}) - P_\text{dia}(t)}{R_4} & \text{if } P_\text{dia}(t) < P_\text{dia}(t_\text{cycle}) \quad \text{(Filling)}
\end{cases}
\label{Equation_11}
\end{equation}

Similar to the systole, Eq.~\ref{Equation_11} can be elegantly expressed using a maximum function:

\begin{equation}
Q_\text{dia}(t) = \max \left(0, \frac{P_\text{dia}(t_\text{cycle}) - P_\text{dia}(t)}{R_4} \right)
\label{Equation_12}
\end{equation}

This formulation ensures that the volume trajectory returns to its baseline state while accurately replicating the IVR phase. The filling phase initiates when $P_\text{dia}(t) < P_\text{dia}(t_\text{cycle})$, intrinsically routing the incoming flow through the respective resistive pathways. The underlying pressure recovery is directly forced by the decay of the active pressure ($\frac{dP_\text{act}(t)}{dt}$) until relaxation is complete, governed by the capacitive node equation:

\begin{equation}
\frac{dP_\text{dia}(t)}{dt} = \frac{dP_\text{act}(t)}{dt} + \frac{1}{C_2} \left(Q_\text{dia}(t)- \frac{P_\text{dia}(t)}{R_3} \right)
\label{Equation_13}
\end{equation}

When the actuator relaxes rapidly and remains at zero for the duration of diastole ($P_\text{act}=0$), Eq.~\ref{Equation_13} reduces to the following:

\begin{equation}
\frac{dP_\text{dia}(t)}{dt} = \frac{1}{C_2} \left[ \max\left(0, \frac{P_\text{dia}(t_\text{cycle}) - P_\text{dia}(t)}{R_4}\right) - \frac{P_\text{dia}(t)}{R_3} \right]
\label{Equation_14}
\end{equation}

This implies that $P_\text{dia}(t)$ follows an exponential recovery, as Eq.~\ref{Equation_14} behaves as a first-order ordinary differential equation.

The location of $P_\text{act}(t)$ in the phase-dependent AWK3 model employs distinct circuit topologies for systole and diastole to accurately reflect the underlying mechanical-to-hydraulic analogies of the active-passive coupling. During systole, the actuation unit actively compresses the ventricular chamber to eject fluid. Consequently, the actuation pressure ($P_{\text{act}}$) acts as an external driving source in series with the proximal resistance ($R_1$). This series configuration physically represents the internal viscoelastic damping of the actuator and the mechanical impedance of the outlet valve, which must be overcome before energy is transmitted to the systemic network. The resulting ventricular pressure ($P_{\text{sys}}$) then drives the flow into the distal resistance ($R_2$) and systemic compliance ($C_1$). Conversely, the diastolic phase involves the rapid relaxation of the actuator, which mechanically pulls the chamber walls to induce expansion. To precisely implement isovolumetric relaxation and subsequent filling, $P_{\text{act}}$ is positioned in series with the chamber's diastolic compliance ($C_2$). This direct structural coupling ensures that the rapid decay of the active force autonomously forces the internal pressure to drop while the volume remains initially constrained. Furthermore, this $C_2 - P_{\text{act}}$ branch is configured in parallel with the refilling resistance ($R_3$). By employing this parallel topology, the model organically transitions between the two diastolic stages: it successfully maintains the isolated pressure drop during isovolumetric relaxation and subsequently allows fluid influx through $R_3$ once the ventricular suction pressure initiates filling.

The physical and fluidic parameters identified using the governing equations are summarized in Table~\ref{Table2}. These values were grounded in physiologically defined lumped-parameter models of the human cardiovascular system \cite{westerhof2009arterial, shi2011review}. Specifically, the systemic parameters ($C_\text{1}$, $C_\text{2}$) reflect the normal physiological ranges of the human arteries. Meanwhile, the proximal and refilling resistances and the viscous resistance ($R_\text{4}$) account for the lumped mechanical impedance of the artificial valves and the inherent viscoelasticity of the soft robotic materials, respectively ($R_\text{1}$, $R_\text{3}$) \cite{polygerinos2015modeling, timms2005complete}. Based on these physiological criteria, the parameters were systematically calibrated to match the behaviors of the MCL.

\begin{table}[t]
    \centering
    \footnotesize
    \renewcommand{\arraystretch}{1.2}
    \caption{\textbf{System parameters for the simulation.} \\ Summary of hydraulic and mechanical parameters calibrated for both SEA-driven and TCPA-driven systems.}
    \label{Table2}
    \begin{tabular*}{\textwidth}{@{\extracolsep{\fill}} l l c c }
        \toprule
        \textbf{Parameter} & \textbf{Description} & \textbf{SEA} & \textbf{TCPAs} \\ 
        \midrule
        $R_\text{1}$ & Systolic proximal resistance ($\text{mmHg}\cdot\text{s/mL}$) & $11$ & $22$ \\ 
        $R_\text{2}$ & Systolic distal resistance ($\text{mmHg}\cdot\text{s/mL}$) & $2.4$ & $1.8$ \\
        $R_\text{3}$ & Diastolic refilling resistance ($\text{mmHg}\cdot\text{s/mL}$) & $6.0  \times 10^1 $ & $1.8$ \\
        $R_\text{4}$ & Viscous damping coefficient ($\text{mmHg}\cdot\text{s/mL}$) & $0.80$ & $14$ \\
        $C_\text{1}$ & Systolic compliance ($\text{mL/mmHg}$) & $0.25$ & $0.55$ \\ 
        $C_\text{2}$ & Diastolic compliance ($\text{mL/mmHg}$) & $0.11$ & $0.45$ \\
        $P_\text{th}$ & Outlet valve threshold pressure ($\text{mmHg}$) & $4.0 \times 10^1$ & $3.0 \times 10^1$ \\
        $k_\text{s}$ & Series spring stiffness ($\text{N/m}$) & $2.5 \times 10^3$ & -- \\
        $k_\text{c}$ & Intrinsic chamber stiffness ($\text{N/m}$) & $2.9 \times 10^3$ & $7.1 \times 10^2$ \\
        $k_\text{p}$ & Unit parallel spring stiffness ($\text{N/m}$) & $1.4 \times 10^2$ & -- \\
        $A_\text{c}$ & Cross-sectional area of ventricular chamber ($\text{m}^2$) & \multicolumn{2}{c}{$1.2 \times 10^{-3}$} \\ 
        \bottomrule
    \end{tabular*}
\end{table}

\subsection{SEA-driven Soft Robotic Ventricle: Experimental Validation}\label{Section3-2}

\begin{figure}[tp!]
    \centering
    \includegraphics[width=0.5\linewidth]{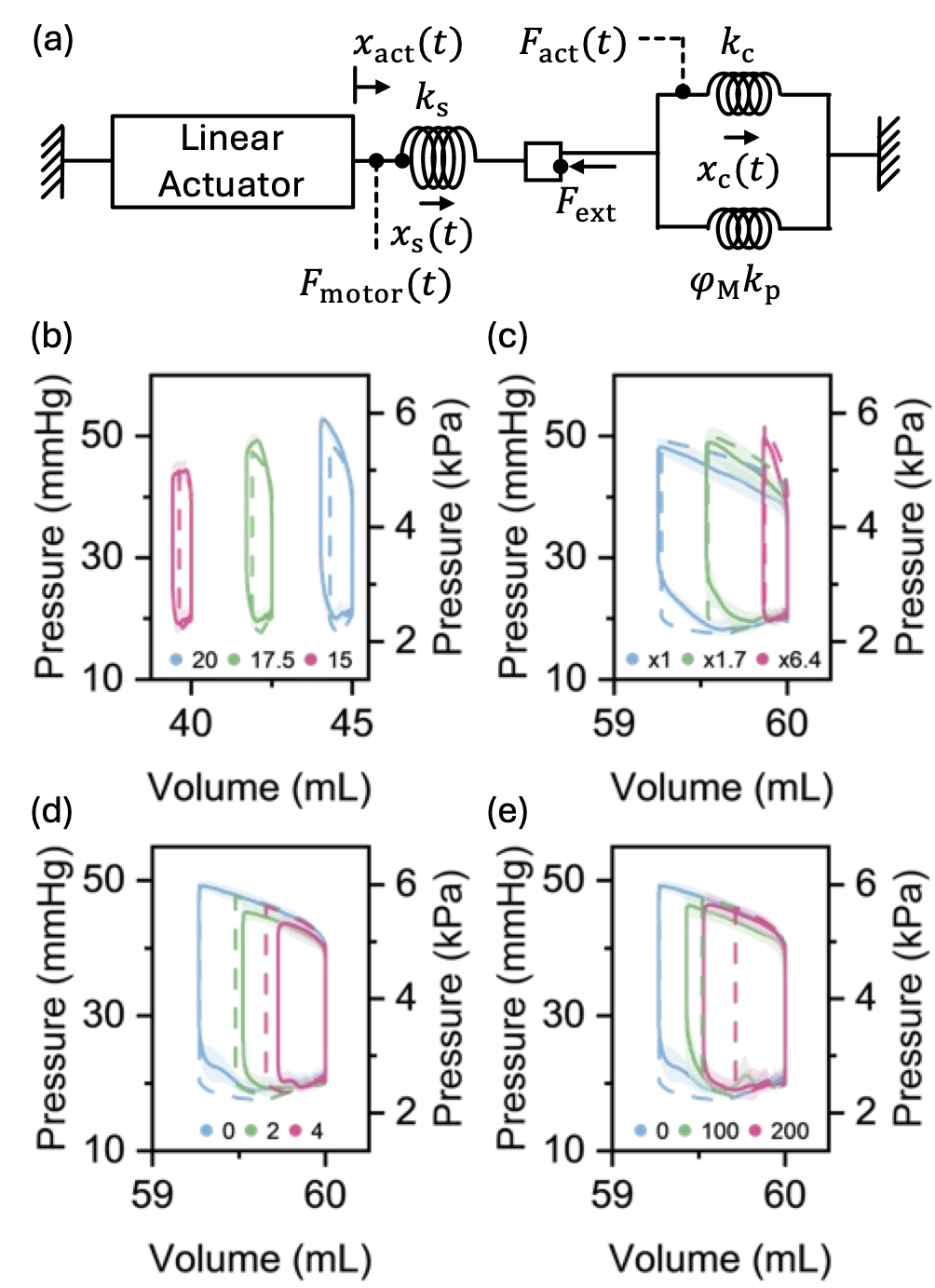}
    \caption{\textbf{Validation of the phase-dependent AWK3 model via an SEA-driven soft robotic ventricle.} (a) Mechanical schematic of the series elastic actuator (SEA) coupled to the soft robotic ventricle, defining displacements ($x_\text{act}(t)$, $x_s(t)$, $x_c(t)$), forces ($F_\text{motor}(t)$, $F_\text{act}(t)$, $F_\text{ext}$), and stiffness parameters of the series spring ($k_s$), parallel springs ($\varphi_\text{M} k_p$), and the chamber ($k_c$). (b--e) Comparison between experimental (dashed lines) and simulation (solid lines) Pressure-Volume (P-V) loops under varying physiological conditions. (b) Preload variations: Adjusting the initial actuator position constrains the end-diastolic volume. (c) Afterload variations ($R_\text{2}$): Increasing the outlet valve resistance elongates the P-V loop vertically, raising peak systolic pressure while reducing stroke volume. The legend values indicate the scaling factors applied to the baseline $R_\text{2}$ parameter detailed in Table~\ref{Table2}. (d--e) Inotropy variations: Modulating the net contractile strength of the system through (d) changing the number of parallel compression springs ($\varphi_\text{M}$), and (e) applying external gravitational loads ($F_\text{ext}$). Each method demonstrates a reduction in net inotropy, accurately reflecting the simultaneous decreases in peak systolic pressure and stroke volume.}
    \label{fig:Figure4}
\end{figure}

To validate the phase-dependent AWK3 model under rigorously controlled conditions, we first employed the SEA-driven configuration. The actuation unit consists of a linear actuator exerting a force of $F_\text{motor}(t)$ (Appendix~\ref{Appendix5}) in series with a spring ($k_\text{s}$) and a ventricular chamber with a stiffness of $k_\text{c}$ (Fig.~\ref{fig:Figure4}a). A set of $\varphi_\text{M}$ parallel compression springs, each with a spring constant $k_\text{p}$, act in parallel with the chamber, and an external load ($F_\text{ext}$) opposes the advance of the actuator.

We first state the kinematics. Assuming the baseline system is grounded at both ends, the
displacement of the linear actuator ($x_\text{act}$) is the sum of the series-spring displacement ($x_\text{s}$) and the chamber displacement ($x_\text{c}$). The actuator travel is limited to $L_\text{max} = 30~\text{mm}$, of which the end-diastolic pre-displacement ($x_\text{ED}$) is already consumed, so the total excursion measured from the fully retracted position obeys:

\begin{equation}
    x_\text{act}(t) + x_\text{ED} = x_\text{s}(t) + x_\text{c}(t) + x_\text{ED} \le L_\text{max}
\label{Equation_15}
\end{equation}

Importantly, $x_\text{ED}$ enters the model only through this kinematic constraint and not as a
stored elastic force. The pre-displacement is imposed while the circuit is open, so the displaced fluid escapes and the chamber relaxes at the new position; because the viscoelastic recovery time scale of the chamber is much longer than the cardiac cycle, it remains effectively at equilibrium throughout diastole and exerts no restoring force at the onset of systole. This is consistent with the end-diastolic baseline pressure, which is identical across all pre-displacement conditions.

Equation~\ref{Equation_15} separates two mechanical regimes. When the inequality is strict,
the actuator is free to advance and the applied force determines the displacements. Once the
equality is reached, the actuator is arrested and the fixed travel $L_\text{max} - x_\text{ED}$ is instead redistributed between $x_\text{s}$ and $x_\text{c}$ in inverse proportion to their
stiffnesses; any surplus motor output is then reacted by the end stop rather than transmitted to
the fluid. Since $F_\text{motor}$ is directly transmitted to the series spring, its displacement
is:

\begin{equation}
    x_\text{s}(t) =
    \begin{cases}
    \dfrac{F_\text{motor}(t)}{k_\text{s}}
        & \text{if} \quad x_\text{act}(t) + x_\text{ED} < L_\text{max} \\[8pt]
    \left(k_\text{c}+\varphi_\text{M}k_\text{p}\right) \frac{L_\text{max} - x_\text{ED}}{k_\text{s} + k_\text{c} + \varphi_\text{M} k_\text{p}}
        & \text{if} \quad x_\text{act}(t) + x_\text{ED} = L_\text{max}
    \end{cases}
\label{Equation_16}
\end{equation}

The external load acts as a mechanical threshold: force is transmitted to the chamber and the
parallel springs only when the force in the series spring exceeds $F_\text{ext}$. Combining this
threshold with the two regimes of Eq.~\ref{Equation_15}, the chamber displacement follows
three regimes:
\begin{equation}
    x_\text{c}(t) =
    \begin{cases}
    0
        & \text{if} \quad F_\text{motor}(t) < F_\text{ext} \\[4pt]
    \dfrac{F_\text{motor}(t) - F_\text{ext}}{k_\text{c} + \varphi_\text{M} k_\text{p}}
        & \text{if} \quad F_\text{motor}(t) \ge F_\text{ext}
          \ \text{and} \ x_\text{act}(t) + x_\text{ED} < L_\text{max} \\
    k_\text{s}\,\dfrac{L_\text{max} - x_\text{ED}}{k_\text{s} + k_\text{c} + \varphi_\text{M} k_\text{p}}
        & \text{if} \quad x_\text{act}(t) + x_\text{ED} = L_\text{max}
    \end{cases}
\label{Equation_17}
\end{equation}

where the first regime reflects the force threshold imposed by $F_\text{ext}$, the second is
force-determined, and the third is kinematically determined by Eq.~\ref{Equation_15}.
Equation~\ref{Equation_17} can be written compactly as:

\begin{equation}
    x_\text{c}(t) = \min\left[
    \max\left(0,\ \frac{F_\text{motor}(t) - F_\text{ext}}{k_\text{c} + \varphi_\text{M} k_\text{p}}\right),\
    k_\text{s}\,\frac{L_\text{max} - x_\text{ED}}{k_\text{s} + k_\text{c} + \varphi_\text{M} k_\text{p}}
    \right]
\label{Equation_18}
\end{equation}

The effective force contributing to the pumping of the ventricular chamber is defined at the $k_\text{c}$ branch, as illustrated in Fig.~\ref{fig:Figure4}a. Combining Eqs.~\ref{Equation_16}--\ref{Equation_18}, the actuation pressure
$P_\text{act}$ is:

\begin{equation}
    P_\text{act, SEA}(t) = \frac{k_\text{c}\,x_\text{c}(t)}{A_\text{c}}
\label{Equation_19}
\end{equation}

Because the second argument of the minimum contains no dependence on $F_\text{motor}$, the peak actuation pressure in the pre-displaced conditions is set by the geometry alone and is independent of the PWM duty cycle, which is consistent with the measurements in
Fig.~\ref{fig:Figure4}b. Eq.~\ref{Equation_19} indicates that the net force transmitted through the series spring ($k_\text{s}x_\text{s}$) is balanced by the repelling forces of the ventricular chamber ($k_\text{c}$) and parallel springs ($k_\text{p}$) sharing the displacement $x_\text{c}(t)$, along with the external load ($F_\text{ext}$). The primary advantage of this SEA baseline lies in its explicit calculation of $P_\text{act, SEA}$. By implementing the series spring as a force sensor, we can precisely isolate and measure the forces acting at the ventricular chamber using its spring constant ($k_\text{s}$) and its displacement ($x_\text{s}$). This capability is valuable for validation of the phase-dependent AWK3 model, since explicitly decoupling these forces is challenging within fully integrated TCPA systems.

Based on this, we systematically investigated three distinct cases using the SEA-driven platform to demonstrate the capability of our reduced-order phase-dependent AWK3 model to replicate physiological hemodynamic responses—commonly described by the Frank-Starling law: (i) preload adjusted via the initial location of the linear actuator; (ii) afterload controlled by the degree of outlet valve opening; and (iii) inotropy modulated by the actuator's duty cycle, $\varphi_\text{M}$, and external loads ($F_\text{ext}$) (Fig.~\ref{fig:Figure4}b--e). A representative motion of the SEA-driven soft pump is shown in Movie S1.

The preload was tuned by varying the initial position of the linear actuator ($x_\text{ED}$) as $15, 17.5$, and $20~\text{mm}$, while the maximum stroke was fixed by the linear actuator ($L_\text{max}$) to 30 mm, thereby constraining the initial end-diastolic volume (EDV) of the ventricular chamber. In accordance with the Frank-Starling mechanism, a more filled chamber prior to systole enabled the ejection of a larger stroke volume (SV) while reaching higher peak pressures, expanding the P-V loop toward the right (Fig.~\ref{fig:Figure4}b). Furthermore, the simulation accurately replicated the phenomenon where all loops abruptly converge at the end of systole. This represents the mechanical end-stop where further compression is physically restricted, demonstrating the model's exceptional fidelity to hardware-specific boundary conditions.

As depicted in Fig.~\ref{fig:Figure4}c, tightening the outlet valve increased $R_\text{2}$ and thereby the afterload, which raised the peak systolic pressure while reducing the stroke volume, elongating the P--V loop vertically and narrowing its width. The simulation results demonstrated high accuracy in capturing these experimental trends. 

Also, inotropy—representing the contractile strength of the ventricle—was modulated through three different physical variables to mimic the varying contractility of the heart. Adding parallel compression springs provided a stronger antagonistic force against chamber compression, effectively reducing the net inotropy of the system (Fig.~\ref{fig:Figure4}d). Similarly, increasing the external antagonistic load yielded a comparable reduction in contractile efficacy (Fig.~\ref{fig:Figure4}e). As $P_\text{act, SEA}$ is negatively affected by parallel springs ($\varphi_\text{M}k_\text{p}$) and $F_\text{ext}$, it is expected that the larger resisting force results in less pressure and volume ejection. Across all parameters, the phase-dependent AWK3 model accurately replicated the dynamic trajectories and non-linear behaviors observed in the SEA-driven soft robotic ventricle.

\subsection{Force Analysis of the TCPA via Isometric Testing}\label{Section3-3}

\begin{figure}[tp!]
    \centering
    \includegraphics[width=1\textwidth]{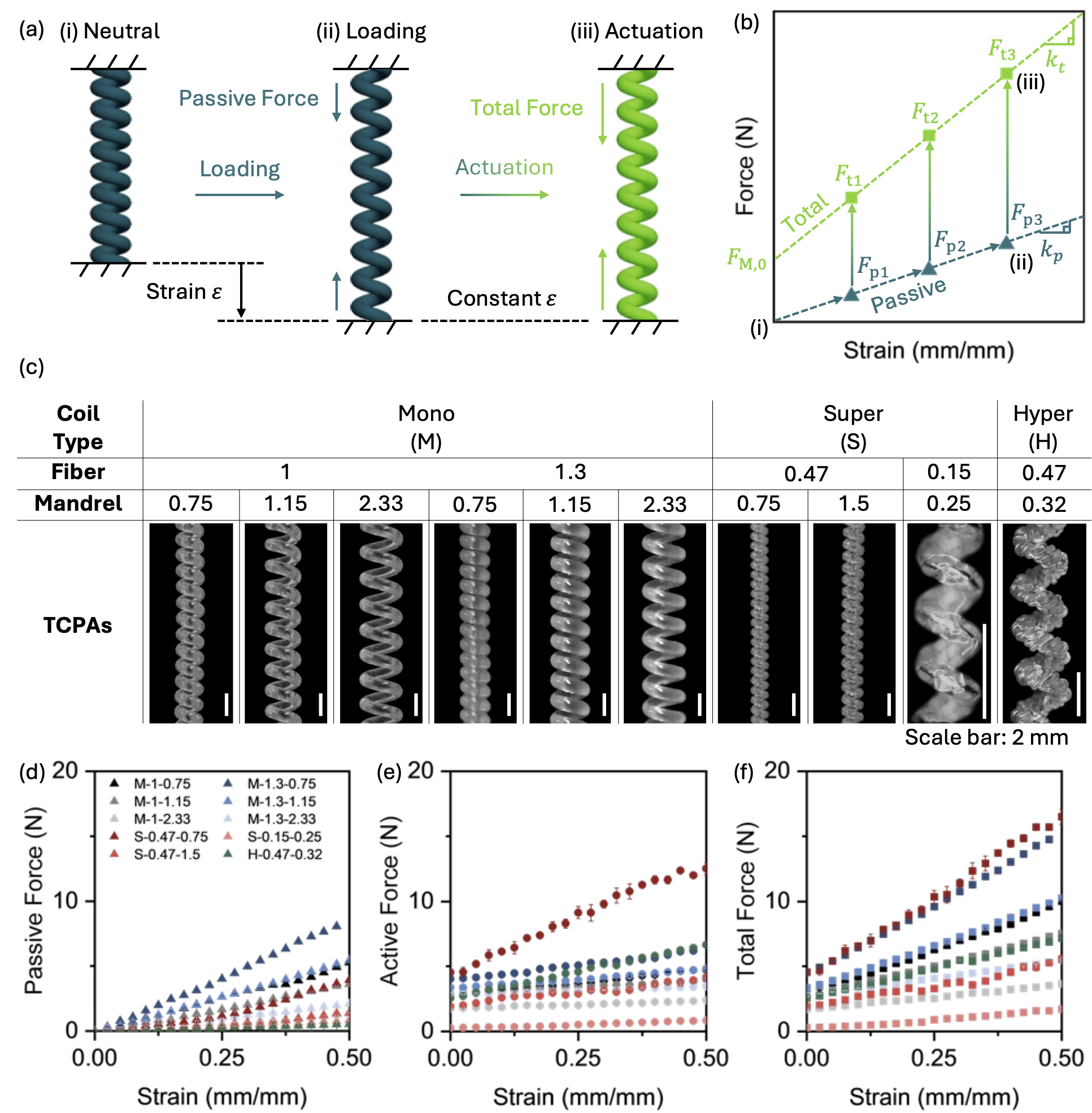}
    \caption{\textbf{Overview of the isometric test.} (a) Schematic of the isometric test procedure. (b) Representative force-strain curves obtained from the isometric test. (c) Optical images of monocoiled, supercoiled, and hypercoiled TCPAs. (d--f) Comparison of passive, active, and total forces across various TCPA configurations.}
    \label{fig:Figure5}
\end{figure}

Fig.~\ref{fig:Figure5}a--b illustrate the experimental protocol for the isometric characterization of the TCPAs. We measured the passive force at room temperature prior to thermal actuation, and the total force at the maximum actuation temperature, under constant strain conditions. This process was repeated across incremental strain steps. The active force, defined as the net mechanical contribution of thermal actuation, was determined by subtracting the passive force from the total measured force. As illustrated in Fig.~\ref{fig:Figure5}b, we derived the passive stiffness ($k_\text{p}$) and total stiffness ($k_\text{t}$) from the slopes of the passive and total force curves, respectively. Additionally, the TCPAs exhibit a non-zero total force even at zero strain, which we define as $F_\text{M,0}$. This formulation implies that by determining the strain of the TCPAs under specific conditions, we can accurately predict the resulting passive and total forces.

We characterized various TCPA geometries, employing a nomenclature defined by the coil type—fiber diameter—mandrel diameter (M: monocoil, S: supercoil, and H: hypercoil). For example, S-0.47-0.75 denotes a supercoiled TCPA fabricated from a 0.47 mm nylon fiber using a 0.75 mm mandrel.

Displacement data were normalized by the initial length to calculate strain, and the resulting force–strain relationships are presented in Fig.~\ref{fig:Figure5}d--f. As demonstrated in Fig.~\ref{fig:Figure5}d, passive forces decreased as the mandrel diameter increased across both monocoiled and supercoiled architectures. Regarding active force generation (Fig.~\ref{fig:Figure5}e), the supercoiled variants exhibited superior performance compared to other geometries, with the S-0.47-0.75 configuration achieving the highest output. Consistent with the trends observed in passive forces, active force generation was inversely correlated with mandrel size. Although hypercoiled muscles generated lower active forces than the optimal supercoiled variant, their performance was comparable to that of standard monocoiled muscles. Finally, the total force profiles (Fig.~\ref{fig:Figure5}f) consistently reflected these trends, confirming the additive relationship between passive and active force components. Further analysis and the implications of these isometric test results are detailed in Appendix~\ref{Appendix2}.

\subsection{TCPA-driven Soft Robotic Ventricle: Modeling and Experimental Validation}\label{Section3-4}

Following the same baseline parameterization, the system was adapted to a fully soft-actuated system using TCPAs (Movie S2, Fig.~\ref{fig:FigureS6}a). Unlike the electromechanical actuation of the SEA configuration, this system converts thermal energy into a pulling force to drive ventricular contraction. While the fluidic network transitions defined by the phase-dependent AWK3 model remains consistent, the active force generation term is reformulated as a state-dependent function to capture the complex thermomechanical behavior of the TCPAs. 

\begin{figure}[tp!]
    \centering
    \includegraphics[width=1\linewidth]{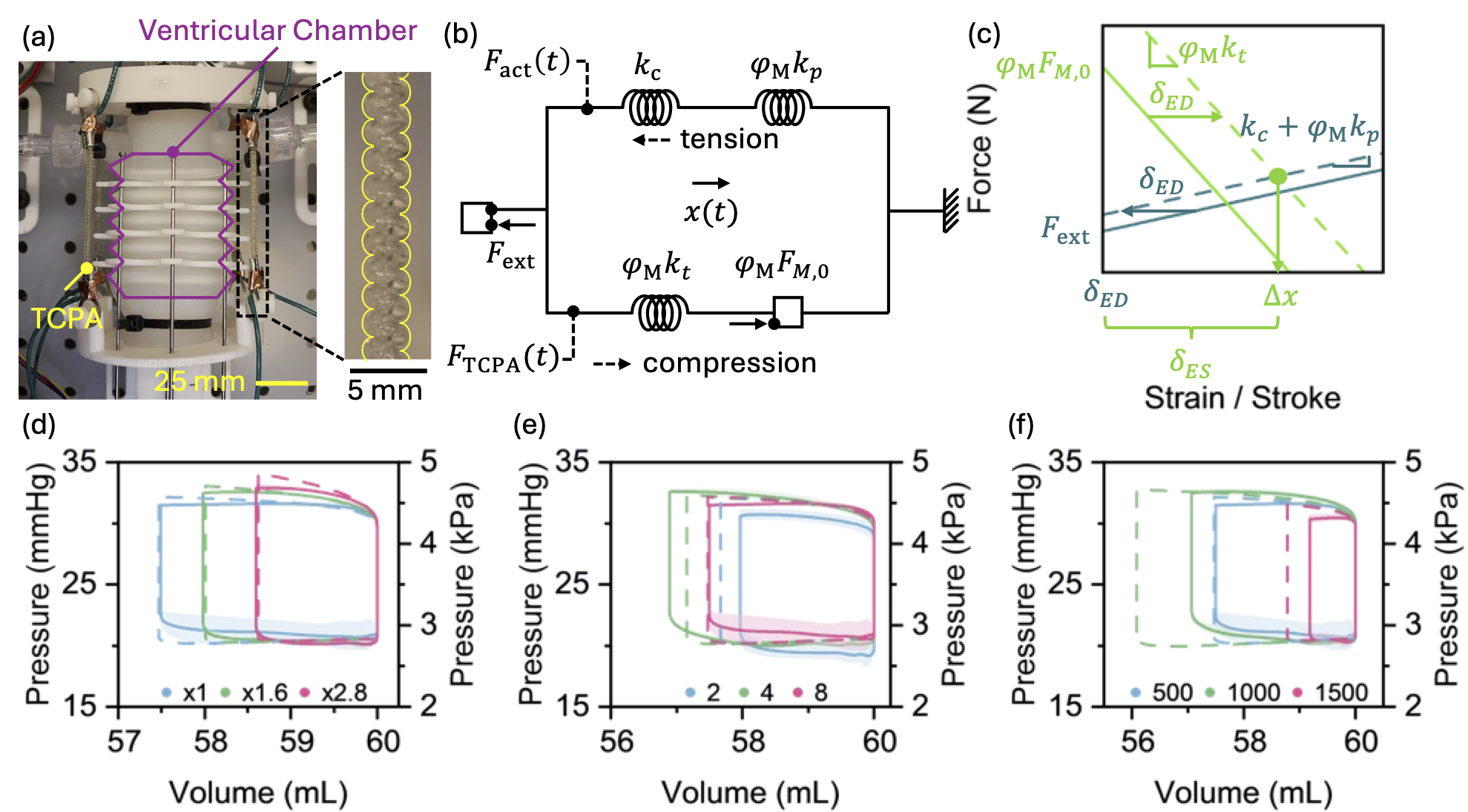}
    \caption{\textbf{Validation of the phase-dependent AWK3 model via a TCPA-driven soft robotic ventricle.} (a) Representative photographs of the ventricular chamber driven by TCPAs. A magnified view of a TCPA is shown on the right. (b) Mechanical reduced order model of the ventricular chamber, as well as the passive and total stiffness of the TCPAs. (c) Force vs strain / stroke curves to describe the antagonistic branches and end-systole displacement ($\delta_\text{ES}$). (d--f) PV loops evaluating the fully soft-actuated system under diverse physiological modulations, comparing experimental results (solid lines) with phase-dependent AWK3 model predictions (dashed lines).  (d) Afterload variations ($R_\text{2}$) where the legend values indicate the scaling factors applied to the baseline $R_\text{2}$ parameter detailed in Table~\ref{Table2}. (e) Structural and contractile scaling by altering the number of integrated TCPA modules ($\varphi_\text{M}$), which simultaneously influences both the inherent passive stiffness and the net active contractility. (f) Performance modulation under varying $F_\text{ext}$ at constant $\varphi_\text{M}=8$.}
    \label{fig:Figure6}
\end{figure}

As described in Fig.~\ref{fig:Figure6}a, the TCPAs with unit number of $\varphi_\text{M}$ are connected in parallel to the ventricular chamber. We model the ventricular structure as two opposing branches (Fig.~\ref{fig:Figure6}b): a passive (upper) branch comprising the ventricular chamber stiffness ($k_\text{c}$) and the passive stiffness of the TCPAs ($\varphi_\text{M}k_\text{p}$), and an active (bottom) branch comprising the post-activation stiffness ($\varphi_\text{M}k_\text{t}$) together with the strain-independent active force ($\varphi_\text{M}F_\text{M,0}$). Both branches are connected in series with $F_\text{ext}$. The passive branch and $F_\text{ext}$ exert force in the direction that elongates the TCPAs, whereas the active branch acts in the opposite direction and tends to contract them. This configuration translates the TCPA-driven soft ventricular system into the antagonistic structure described in Appendix~\ref{Appendix4}, in which the intersection of the passive curve and the flipped active curve predicts the stroke assuming the quasi-static actuation \cite{wang2026characterization}.

Fig.~\ref{fig:Figure6}c illustrates that unlike the simple antagonistic structure in Appendix~\ref{Appendix4}, all passive and total force components must be considered. Here, $\delta$ denotes a deformation measured from the free length of the chamber--TCPA assembly, that is, the configuration at which neither the external load nor the activation is applied. Both $\delta_\text{ED}$ and $\delta_\text{ES}$ are defined as magnitudes, and their directions are stated explicitly where they are introduced.

Because the force generated by a TCPA depends on its strain (Section~\ref{Section3-3}), the operating point must first be located. Prior to activation the external load is carried jointly by the chamber and the passive stiffness of the TCPAs, which are mechanically in parallel. The assembly is therefore elongated from its free length by

\begin{equation}
\delta_\text{ED} = \frac{F_\text{ext}}{\varphi_\text{M}k_\text{p}+k_\text{c}}
\label{Equation_20}
\end{equation}

which defines the end-diastolic state. Upon activation the two branches oppose each other, and the assembly settles where they balance. Measuring the compression from the free length, this equilibrium reads

\begin{equation}
\underbrace{\left(k_\text{c}+\varphi_\text{M}k_\text{p}\right)\delta_\text{ES}+F_\text{ext}}_{\text{passive branch and external load}} = \underbrace{\varphi_\text{M}F_\text{M,0}-\varphi_\text{M}k_\text{t}\,\delta_\text{ES}}_{\text{active branch}}
\label{Equation_21}
\end{equation}

so that the end-systolic compression is

\begin{equation}
\delta_\text{ES} = \frac{\varphi_\text{M}F_\text{M,0} - F_\text{ext}}{k_\text{c}+\varphi_\text{M}\left(k_\text{p}+k_\text{t}\right)}
\label{Equation_22}
\end{equation}

Note that $k_\text{p}$ is retained in the denominator: activation superposes $\varphi_\text{M}F_\text{M,0}$ and $\varphi_\text{M}k_\text{t}$ on the TCPAs but does not remove their passive elasticity, in analogy with the parallel elastic element of the Hill muscle model. Equation~\ref{Equation_22} also delimits the operating range of the model: contraction occurs only when $F_\text{ext}<\varphi_\text{M}F_\text{M,0}$, that is, when the TCPA bundle can overcome the applied load.

Since $\delta_\text{ED}$ is an elongation and $\delta_\text{ES}$ a compression, the two are directed oppositely with respect to the free length. Over one systole the assembly must first recover the elongation imposed by $F_\text{ext}$ and only then compress the chamber, so the stroke is the sum of the two magnitudes:

\begin{equation}
\Delta x = \delta_\text{ED} + \delta_\text{ES} = \frac{F_\text{ext}}{k_\text{c}+\varphi_\text{M}k_\text{p}} + \frac{\varphi_\text{M}F_\text{M,0} - F_\text{ext}}{k_\text{c}+\varphi_\text{M}\left(k_\text{p}+k_\text{t}\right)}
\label{Equation_23}
\end{equation}

This is the interval marked in Fig.~\ref{fig:Figure6}c, where the passive curve is offset by $+\delta_\text{ED}$ and the flipped active curve by $-\delta_\text{ED}$, and their intersection locates the end-systolic operating point. The opposing directions also explain the opposite signs of $F_\text{ext}$ in the two terms of Eq.~\ref{Equation_23}: a larger load elongates the assembly further, which aids filling, but simultaneously reduces the contraction the TCPAs can produce against it. Taking the force at the $k_\text{c}$ branch as the effective pumping force, the activation pressure is

\begin{equation}
P_\text{act, TCPA}(t) = \frac{k_\text{c}\,\Delta x}{A_\text{c}}\,T(t) = \frac{k_\text{c}}{A_\text{c}}\left[\frac{F_\text{ext}}{k_\text{c}+\varphi_\text{M}k_\text{p}} + \frac{\varphi_\text{M}F_\text{M,0} - F_\text{ext}}{k_\text{c}+\varphi_\text{M}\left(k_\text{p}+k_\text{t}\right)}\right]T(t)
\label{Equation_24}
\end{equation}

where $T(t)$ is the normalized thermal response of the TCPAs, which rises with the heating time constant $\tau_\text{h}$ during systole and decays with the cooling time constant $\tau_\text{c}$ during diastole (Appendix~\ref{Appendix6}). Unlike the SEA-driven configuration, the actuator here is not kinematically constrained, so $P_\text{act, TCPA}$ scales directly with the active force $\varphi_\text{M}F_\text{M,0}$ and hence with the input energy supplied to the TCPAs.

The dynamic thermal actuation is governed by $T(t)$, a time-varying temperature modulation factor that smoothly scales the maximum isometric force generation over the systolic duration $t_{\text{sys}}$ (Appendix~\ref{Appendix6}).

To experimentally evaluate the adaptability of the phase-dependent AWK3 framework to this fully soft-actuated system, a representative TCPA (S-0.47-0.75) was employed as the actuator unit. This soft-actuated MCL was designed to dynamically replicate physiological hemodynamics, including the Frank-Starling mechanism, by systematically modulating key experimental parameters. Specifically, we varied: (i) closing of outlet valve, (ii) the number of active TCPAs, (ii) the electrical energy input, and (iii) various $F_\text{ext}$ representing the external diastolic load. By measuring the resultant pressure-volume (P-V) loops across these diverse conditions, the predictive accuracy of the phase-dependent AWK3 model was comprehensively validated within a fully soft robotic environment.

When the outlet valve was constricted to increase the peripheral resistance ($R_\text{2}$), the system exhibited a concomitant rise in pressure and reduction in stroke volume, consistent with the SEA-driven baseline behavior (Fig.~\ref{fig:Figure6}d). The non-monotonic relationship observed in the mechanical work output reflects the complex interaction between mechanical impedance and actuator efficiency (Fig.~\ref{fig:Figure6}e--f). As demonstrated by the isometric characterization, insufficient load per unit fails to adequately pre-stretch the TCPAs, thereby limiting the effective addition of active stress. This inefficiency manifests as a reduced stroke volume and diminished mechanical work at lower module counts as reflected in the loop areas. Conversely, as the number of modules increases beyond the optimal range, the system encounters excessive passive resistance. This phenomenon, which can be explained by mechanical impedance, necessitates higher energy input to achieve the same deformation, thereby reducing overall work output.

\section{Discussion}

\subsection{Utility of the phase-dependent AWK3 model}
The phase-dependent AWK3 model developed in this work enables the design and analysis of the circulatory soft pump driven by a soft actuator. This is a paradigm shift from the existing approaches of designing and constructing a circulation system by analyzing the open-loop fluidic equations, while relying mostly on trial-and-error to find a suitable driving linear actuator. This is enabled by embedding the state-dependent active and passive stiffness of the actuators which were derived from isometric testing directly into the hydraulic governing equations. The framework, which builds on the phase-dependent AWK3 model, is used to select a suitable actuator along with designing the circulatory fluidic system, and simultaneously co-predicts both pressure and volume of each stroke. It calculates the instantaneous dynamic equilibrium between the actuator internal stress generation and the external fluidic impedance. This advancement enables the rigorous mathematical replication of physiological hemodynamics, including the Frank-Starling mechanism, without relying on arbitrary, predefined input waveforms.

Our work addressed the fundamental limitations of the WK model. Historically, lumped-parameter WK models have been successful in representing systemic arterial impedance \cite{westerhof2009arterial, christ2020hydraulic}. However, they have conventionally been utilized as passive, "open-loop" or "uncoupled" boundary conditions using either pressure or flow rate to predict each other \cite{westerhof2009arterial, demeersseman2025activation, rocchi2024patient}. In this classical approach, the model requires either a predefined pressure waveform to calculate volume, or a predefined flow rate to calculate pressure. It acts merely as a passive fluidic network, entirely decoupled from the actual force-generation mechanics of the contractile elements driving the flow. This uncoupled nature makes it virtually impossible to accurately simulate the critical isovolumetric phases, where actuator force rapidly builds up against closed valves without any change in volume.

We validated the model using two different types of actuators to show its general applicability. Rather than using a stiff electromechanical actuator on its own, we added a soft spring in series to pave the way to validation using the soft TCPA. SEAs were originally developed to provide intrinsic compliance to robotic systems, enabling safe interactions and energy storage akin to natural tendons and muscles \cite{pratt1995series}. Despite the widespread and successful adoption of SEA in musculoskeletal biomechanics and legged locomotion, their integration into soft robotic heart simulators has been notably absent. Previous soft robotic hearts have used more complex soft actuators, such as pneumatic soft actuators, that are highly nonlinear and difficult to parameterize \cite{roche2017soft, rosalia2022soft, guex2021increased, ueda2025soft}. By introducing an SEA-driven baseline in our study, we effectively create a reference case for a soft actuator driven soft pump. The SEA platform allows us to explicitly parameterize and validate the active-passive mechanical coupling within the cardiac cycle—specifically isolating the interplay between the contractile motor, the series compliance, and the pumping chamber before transitioning to the complex thermomechanical behaviors of TCPAs.

\subsection{TCPA Isometric Test Insights}

The isometric test quantifies the two independent components of force—passive and total forces—that contribute to the overall performance of the muscles \cite{zhang2019modeling}. This capability provides a deeper understanding of the underlying mechanisms and dynamics of TCPAs, especially compared to other conventional testing techniques, such as tensile and isobaric tests. When subjected to high temperatures, TCPAs made of nylon, a thermoelastic material, undergo softening, resulting in real-time changes in their properties \cite{chen2024effect}. However, the tensile test, typically conducted at room temperature, fails to consider these temperature-dependent variations in stiffness. Also, the isobaric test focuses primarily on quantifying the stroke capabilities of TCPAs under specific applied forces. Although this provides valuable information regarding the extent of stroke, it is unable to isolate and analyze the individual passive and active force contributions. Understanding the separate contributions of passive and active forces is absolutely essential when TCPAs are integrated into complex systems like soft robotic hearts, where each force term organically interacts with external passive elastomeric chambers. By controlling the applied strain and precisely measuring the resulting forces, the isometric test enables the optimization of TCPA design for phase-dependent cardiac applications.

The isometric test presents advantages to enable systematic design of soft pumps driven by TCPAs. However, while the isometric test provides a controlled environment to evaluate TCPA performance, its artificial suppression of stress relaxation must be carefully interpreted. Given that TCPAs are exposed to temperatures beyond the glass transition temperature of the raw material during testing, stress relaxation is not avoidable \cite{rafie2018characterizing}. By physically constraining the actuator at a constant strain, the isometric test reduces macroscopic deformation of TCPAs. Although this ability is particularly advantageous in that it enables direct assessment and comparison between different TCPAs, it inherently masks the continuous performance degradation caused by stress relaxation. Consequently, this constraint yields an idealized measurement that can hide the actuator's hysteresis  which plays an important role in dynamic, real-world systems like a soft robotic heart, where constrained time-dependent deformation inevitably compromises the effective stroke. Therefore, while isometric data is essential for isolating intrinsic contractile properties, it must be understood as an upper bound of performance rather than a direct reflection of dynamic operational capacity.

The capability of TCPAs to generate higher force under isometric actuation enables switching between isometric and isotonic conditions \cite{zhang2019modeling, roberts1997muscular}. Isometric contraction is found not only in the musculoskeletal system \cite{roberts1997muscular, roberts2010series}, but also in cyclic cardiac muscles \cite{hall2020guyton, burkhoff1986ventricular}. For example, the natural heart ejects blood from the left ventricle by dramatically increasing ventricular pressure during a short period of strictly constrained volume. This natural behavior highlights the direct relevance of isometric testing for developing soft robotic hearts. To accurately replicate the IVC phase—where ventricular pressure must rise rapidly against closed valves—the soft artificial ventricle must generate substantial force without volumetric strain. TCPAs, which build up significant active internal stress under isometric constraints, flawlessly emulate this biological energy-storage mechanism prior to systolic ejection.

\subsection{Quantitative Design Framework for Soft Artificial Hearts}
The development of bio-inspired soft robotic hearts has seen remarkable progress in recent years, with numerous platforms successfully replicating the complex anatomy and macroscopic kinematic motions of the native myocardium \cite{park2024biorobotic, roche2017soft, rosalia2022soft, davies2026soft}. However, a persistent challenge in the field has been the heavy reliance on heuristic, trial-and-error methodologies to achieve target hemodynamic outputs. In previous studies, matching the physiological P-V loop of a natural heart required iterative physical prototyping—manually adjusting elastomer thicknesses, empirically altering actuation pressures, or continuously tweaking control timings until the desired performance was observed \cite{rosalia2024modulating, guex2021increased}. While computational tools like FEA are frequently employed \cite{zhang2019modeling, demeersseman2025activation, park2022computational}, they function primarily as descriptive evaluative tools rather than generative design frameworks. They lack the inverse capability to predict the necessary actuator configurations required to overcome specific systemic loads.

This study presents a critical breakthrough by establishing a deterministic, forward-design strategy for soft artificial hearts. By systematically integrating the muscle-level isometric characterization of TCPAs with the system-level phase-dependent AWK3 model, we completely bypass the need for arbitrary trial-and-error tuning. Our framework mathematically links the foundational force-strain capabilities of individual soft artificial muscles directly to the global macroscopic pumping performance. Consequently, critical cardiac parameters—such as inotropy, preload, and afterload—can be analytically predicted and finely tuned strictly based on the physical properties of the integrated TCPAs and the hydraulic network. 

This systematic approach dramatically accelerates the development cycle and enhances the reliability of soft robotic cardiovascular systems. It allows researchers and engineers to computationally "prescribe" the precise quantity, mechanical stiffness, and thermal actuation profiles of TCPAs required to support a specific patient's systemic resistance long before any physical assembly occurs. Ultimately, shifting from empirical observation to rigorous mathematical prediction not only enhances the biological fidelity of in vitro mock circulatory loops, but also establishes a robust design protocol for the next generation of implantable, soft robotic ventricular assist devices \cite{weymann2023artificial, kongahage2021high}.

\subsection{Limitations and Future Perspectives}
As with many lumped-parameter approaches, these computationally efficient and accessible models do not capture complex three-dimensional spatial deformation dynamics that occur in natural myocardial tissues \cite{formaggia2010cardiovascular, genet2014distribution}. Second, the framework currently neglects inertial effects—both fluidic and structural—assuming quasi-static operation within the cardiac cycle. While this approximation is valid for the low-frequency hemodynamic behaviors presented in this study, rapid dynamic transitions in future high-rate applications may require the inclusion of inertial terms to fully capture the system's transient response \cite{stergiopulos1999total, courtois1988transmitral}. Third, our MCL setup uses a basic hydraulic network with linear compliance, resistances and valves. Incorporating more complex arterial impedances, such as non-linear compliance or wave reflection effects, would provide a more comprehensive validation of the model's adaptability in diverse pathological conditions. Despite these limitations, the phase-dependent AWK3 framework serves as a versatile blueprint for the next generation of implantable, soft robotic ventricular assist devices, paving the way for patient-specific hemodynamic optimization.

\section{Conclusion}
This study presents a design framework for soft robotic cardiovascular systems using measured actuator characteristics. The framework is made possible by combining (i) a mathematical lumped-parameter model for the flow circuit in the various systole and diastole phases, connected to a model for the forcing or pressure generation of the actuator; and (ii) actuator testing protocol to identify their active and passive force components. Using these two pillars successfully enables the replication of isovolumetric cardiac mechanics without trial and error-based experimental tuning of the fluidic network and the actuator. The muscle-level compliance is mathematically integrated into the novel phase-dependent AWK3 model. Unlike traditional open-loop formulations, the phase-dependent AWK3 framework autonomously co-predicts both pressure and volume by directly mapping the active-passive mechanical coupling to the macroscopic hydrodynamic boundaries. Extensive validations using both an SEA-driven baseline and soft TCPA-actuated MCL demonstrated the high fidelity of the model in replicating physiological hemodynamics. The framework successfully reproduced the Frank-Starling mechanism across diverse conditions of inotropy, preload, and afterload. Ultimately, this study provides a useful approach for soft robotic pump development transitioning from heuristic trial-and-error to deterministic computational design. This predictive blueprint could be used for the next generation of biomimetic cardiac simulators and patient-specific soft ventricular assist devices.

\backmatter


\section*{Acknowledgements}
This work was supported by a JUMP ARCHES grant.

\section*{Author contributions}
J.K. and S.T. conceived the idea and designed the research. J.K., Q.W., L.C., S.T., and S.H.K. performed the experiments and data analysis. J.K. and S.T. wrote the paper. S.T. supervised the project and reviewed the manuscript. All authors contributed to the writing and editing of the manuscript.

\section*{Competing interests}
The authors declare no competing interests.

\section*{Data availability}
The data that support the findings of this study are available from the corresponding author upon reasonable request.



\bibliography{sn-bibliography}

@article{wang2026characterization,
  title={Characterization and Selection of Contractile Actuators for Soft Robotics Operating in the Quasistatic Regime},
  author={Wang, Qiong and Cheng, Liuyang and Kim, Jeongmin and Tsai, Samuel and Roach, Devin and Tawfick, Sameh},
  journal={Advanced Intelligent Systems},
  pages={e70490},
  year={2026},
  publisher={Wiley Online Library}
}

@article{rosalia2024modulating,
  title={Modulating cardiac hemodynamics using tunable soft robotic sleeves in a porcine model of HFpEF physiology for device testing applications},
  author={Rosalia, Luca and Ozturk, Caglar and Wang, Sophie X and Quevedo-Moreno, Diego and Saeed, Mossab Y and Mauskapf, Adam and Roche, Ellen T},
  journal={Advanced Functional Materials},
  volume={34},
  number={8},
  pages={2310085},
  year={2024},
  publisher={Wiley Online Library}
}

@article{park2022computational,
  title={Computational design of a soft robotic myocardium for biomimetic motion and function},
  author={Park, Clara and Ozturk, Caglar and Roche, Ellen T},
  journal={Advanced Functional Materials},
  volume={32},
  number={40},
  pages={2206734},
  year={2022},
  publisher={Wiley Online Library}
}

@article{ueda2025soft,
  title={Soft Robotic Heart Formed with a Myocardial Band for Cardiac Functions},
  author={Ueda, Daiki and Suzumori, Koichi and Nabae, Hiroyuki and Ishikawa, Yuta and Oda, Teiji},
  journal={Soft Robotics},
  volume={12},
  number={4},
  pages={488--497},
  year={2025},
  publisher={SAGE Publications Sage CA: Los Angeles, CA}
}

@article{guex2021increased,
  title={Increased longevity and pumping performance of an injection molded soft total artificial heart},
  author={Guex, Leonard G and Jones, Lewis S and Kohll, A Xavier and Walker, Roland and Meboldt, Mirko and Falk, Volkmar and Schmid Daners, Marianne and Stark, Wendelin J},
  journal={Soft robotics},
  volume={8},
  number={5},
  pages={588--593},
  year={2021},
  publisher={SAGE Publications Sage CA: Los Angeles, CA}
}

@article{davies2026soft,
  title={A Soft Robotic Model for Simulating Heart Valve Disease and Cardiac Interventions},
  author={Davies, James and Nicotra, Emanuele and Zhu, Kefan and Nguyen, Chi Cong and Sharma, Bibhu and Ji, Adrienne and Phan, Phuoc Thien and Wan, Jingjing and Pruscino, Patrick and Truong, Hermione and others},
  journal={Advanced Science},
  pages={e16667},
  year={2026},
  publisher={Wiley Online Library}
}

@article{rosalia2022soft,
  title={A soft robotic sleeve mimicking the haemodynamics and biomechanics of left ventricular pressure overload and aortic stenosis},
  author={Rosalia, Luca and Ozturk, Caglar and Coll-Font, Jaume and Fan, Yiling and Nagata, Yasufumi and Singh, Manisha and Goswami, Debkalpa and Mauskapf, Adam and Chen, Shi and Eder, Robert A and others},
  journal={Nature biomedical engineering},
  volume={6},
  number={10},
  pages={1134--1147},
  year={2022},
  publisher={Nature Publishing Group UK London}
}

@article{kongahage2021high,
  title={High performance artificial muscles to engineer a ventricular cardiac assist device and future perspectives of a cardiac sleeve},
  author={Kongahage, Dharshika and Ruhparwar, Arjang and Foroughi, Javad},
  journal={Advanced Materials Technologies},
  volume={6},
  number={5},
  pages={2000894},
  year={2021},
  publisher={Wiley Online Library}
}

@inproceedings{pratt1995series,
  title={Series elastic actuators},
  author={Pratt, Gill A and Williamson, Matthew M},
  booktitle={Proceedings 1995 IEEE/RSJ international conference on intelligent robots and systems. Human robot interaction and cooperative robots},
  volume={1},
  pages={399--406},
  year={1995},
  organization={IEEE}
}

@article{almond2013berlin,
  title={Berlin Heart EXCOR pediatric ventricular assist device for bridge to heart transplantation in US children},
  author={Almond, Christopher S and Morales, David L and Blackstone, Eugene H and Turrentine, Mark W and Imamura, Michiaki and Massicotte, M Patricia and Jordan, Lori C and Devaney, Eric J and Ravishankar, Chitra and Kanter, Kirk R and others},
  journal={Circulation},
  volume={127},
  number={16},
  pages={1702--1711},
  year={2013},
  publisher={Lippincott Williams \& Wilkins Hagerstown, MD}
}

@article{copeland2012experience,
  title={Experience with more than 100 total artificial heart implants},
  author={Copeland, Jack G and Copeland, Hannah and Gustafson, Monica and Mineburg, Nicole and Covington, Diane and Smith, Richard G and Friedman, Mark},
  journal={The Journal of thoracic and cardiovascular surgery},
  volume={143},
  number={3},
  pages={727--734},
  year={2012},
  publisher={Elsevier}
}

@article{arfaee2025soft,
  title={A soft robotic total artificial hybrid heart},
  author={Arfaee, Maziar and Vis, Annemijn and Bartels, Paul AA and Van Laake, Lucas C and Lorenzon, Lucrezia and Ibrahim, Dina M and Zrinscak, Debora and Smits, Anthal IPM and Henseler, Andreas and Cianchetti, Matteo and others},
  journal={Nature Communications},
  volume={16},
  number={1},
  pages={5146},
  year={2025},
  publisher={Nature Publishing Group UK London}
}

@article{linke2008sense,
  title={Sense and stretchability: the role of titin and titin-associated proteins in myocardial stress-sensing and mechanical dysfunction},
  author={Linke, Wolfgang A},
  journal={Cardiovascular research},
  volume={77},
  number={4},
  pages={637--648},
  year={2008},
  publisher={Oxford University Press}
}

@article{granzier2004giant,
  title={The giant protein titin: a major player in myocardial mechanics, signaling, and disease},
  author={Granzier, Henk L and Labeit, Siegfried},
  journal={Circulation research},
  volume={94},
  number={3},
  pages={284--295},
  year={2004},
  publisher={Lippincott Williams \& Wilkins}
}

@article{ott2025impact,
  title={Impact of complications on survival outcomes in different temporary mechanical circulatory support techniques: a large retrospective cohort study of cardiac surgical and nonsurgical patients},
  author={Ott, Sascha and Germinario, Lorenzo and M{\"u}ller-Wirtz, Lukas M and Nersesian, Gaik and Hennig, Felix and Hommel, Matthias and Ruetzler, Kurt and Stoppe, Christian and Vandenbriele, Christoph and Schoenrath, Felix and others},
  journal={The Journal of Heart and Lung Transplantation},
  volume={44},
  number={6},
  pages={880--891},
  year={2025},
  publisher={Elsevier}
}

@article{sagawa1990translation,
  title={Translation of Otto frank's paper" Die Grundform des arteriellen Pulses" zeitschrift f{\"u}r biologie 37: 483-526 (1899)},
  author={Sagawa, Kiichi and Lie, Reidar K and Schaefer, Jochen},
  journal={Journal of molecular and cellular cardiology},
  volume={22},
  number={3},
  pages={253--254},
  year={1990}
}

@article{burkhoff1986ventricular,
  title={Ventricular efficiency predicted by an analytical model},
  author={Burkhoff, Daniel and Sagawa, KIICHI},
  journal={American Journal of Physiology-Regulatory, Integrative and Comparative Physiology},
  volume={250},
  number={6},
  pages={R1021--R1027},
  year={1986},
  publisher={American Physiological Society Bethesda, MD}
}

@article{westerhof2009arterial,
  title={The arterial windkessel},
  author={Westerhof, Nico and Lankhaar, Jan-Willem and Westerhof, Berend E},
  journal={Medical \& biological engineering \& computing},
  volume={47},
  number={2},
  pages={131--141},
  year={2009},
  publisher={Springer}
}

@book{hall2020guyton,
  title={Guyton and Hall Textbook of Medical Physiology E-Book: Guyton and Hall Textbook of Medical Physiology E-Book},
  author={Hall, John E and Hall, Michael E},
  year={2020},
  publisher={Elsevier Health Sciences}
}

@article{roberts2010series,
  title={The series-elastic shock absorber: tendons attenuate muscle power during eccentric actions},
  author={Roberts, Thomas J and Azizi, Emanuel},
  journal={Journal of Applied Physiology},
  volume={109},
  number={2},
  pages={396--404},
  year={2010},
  publisher={American Physiological Society Bethesda, MD}
}

@article{roberts1997muscular,
  title={Muscular force in running turkeys: the economy of minimizing work},
  author={Roberts, Thomas J and Marsh, Richard L and Weyand, Peter G and Taylor, C Richard},
  journal={Science},
  volume={275},
  number={5303},
  pages={1113--1115},
  year={1997},
  publisher={American Association for the Advancement of Science}
}

@article{wang2024mechanics,
  title={The mechanics and physics of twisted and coiled polymer actuators},
  author={Wang, Qiong and Ghrayeb, Anan and Kim, SeongHyeon and Cheng, Liuyang and Tawfick, Sameh},
  journal={International Journal of Mechanical Sciences},
  volume={280},
  pages={109440},
  year={2024},
  publisher={Elsevier}
}

@article{yang2016top,
  title={A top-down multi-scale modeling for actuation response of polymeric artificial muscles},
  author={Yang, Qianxi and Li, Guoqiang},
  journal={Journal of the Mechanics and Physics of Solids},
  volume={92},
  pages={237--259},
  year={2016},
  publisher={Elsevier}
}

@article{sharafi2015multiscale,
  title={A multiscale approach for modeling actuation response of polymeric artificial muscles},
  author={Sharafi, Soodabeh and Li, Guoqiang},
  journal={Soft matter},
  volume={11},
  number={19},
  pages={3833--3843},
  year={2015},
  publisher={Royal Society of Chemistry}
}

@article{rosalia2023pneumatic,
  title={Soft robotic patient-specific hydrodynamic model of aortic stenosis and ventricular remodeling},
  author={Rosalia, Luca and Ozturk, Caglar and Goswami, Debkalpa and Bonnemain, Jean and Wang, Sophie X and Bonner, Benjamin and Weaver, James C and Puri, Rishi and Kapadia, Samir and Nguyen, Christopher T and others},
  journal={Science robotics},
  volume={8},
  number={75},
  pages={eade2184},
  year={2023},
  publisher={American Association for the Advancement of Science}
}

@article{tsai2025high,
  title={High cycle performance of twisted and coiled polymer actuators},
  author={Tsai, Samuel and Wang, Qiong and Hur, Ohnyoung and Bartlett, Michael D and King, William P and Tawfick, Sameh},
  journal={Sensors and Actuators A: Physical},
  volume={381},
  pages={116041},
  year={2025},
  publisher={Elsevier}
}

@article{peters2018effect,
  title={Effect of chain length dispersity on the mobility of entangled polymers},
  author={Peters, Brandon L and Salerno, K Michael and Ge, Ting and Perahia, Dvora and Grest, Gary S},
  journal={Physical review letters},
  volume={121},
  number={5},
  pages={057802},
  year={2018},
  publisher={APS}
}

@article{hu2024artificial,
  title={Artificial muscles based on coiled conductive polymer yarns},
  author={Hu, Hongwei and Zhang, Shengtao and Zhang, Mengyang and Xu, Jiawei and Salim, Teddy and Li, Yan and Hu, Xinghao and Zhang, Zhongqiang and Cheng, Guanggui and Yuan, Ningyi and others},
  journal={Advanced Functional Materials},
  volume={34},
  number={33},
  pages={2401685},
  year={2024},
  publisher={Wiley Online Library}
}

@article{witham2024coil,
  title={Coil Formation and Biomimetic Performance Characterization of Twisted Coiled Polymer Artificial Muscles},
  author={Witham, Nicholas S and Mersch, Johannes and Selzer, Lukas and Reiche, Christopher F and Solzbacher, Florian},
  journal={Advanced Intelligent Systems},
  pages={2400334},
  year={2024},
  publisher={Wiley Online Library}
}

@article{chen2024effect,
  title={Effect of temperature softening on the actuation performance of twisted and coiled polymer muscles},
  author={Chen, Yaping and Hu, Jiongjiong and Xie, Yuyang and Liu, Lei and Liu, Dabiao},
  journal={Sensors and Actuators A: Physical},
  volume={374},
  pages={115444},
  year={2024},
  publisher={Elsevier}
}

@article{zhang2024compound,
  title={A compound twisted and coiled actuators with payload-insensitive untwisting characteristics},
  author={Zhang, Hao and Yang, Guilin and Shen, Wenjun and Zhang, Haohao and Zheng, Tianjiang and Zhang, Chi and Chen, Tao},
  journal={Sensors and Actuators A: Physical},
  volume={374},
  pages={115407},
  year={2024},
  publisher={Elsevier}
}

@article{rocchi2024patient,
  title={A patient-specific echogenic soft robotic left ventricle embedded into a closed-loop cardiovascular simulator for advanced device testing},
  author={Rocchi, Maria and Papangelopoulou, Konstantina and Ingram, Marcus and Bekhuis, Youri and Claessen, Guido and Claus, Piet and D'hooge, Jan and Donker, Dirk W and Meyns, Bart and Fresiello, Libera},
  journal={APL bioengineering},
  volume={8},
  number={2},
  year={2024},
  publisher={AIP Publishing}
}

@article{demeersseman2025activation,
  title={Activation of a Soft Robotic Left Ventricular Phantom Embedded in a Closed-Loop Cardiovascular Simulator: A Computational and Experimental Analysis},
  author={Demeersseman, Nele and Rocchi, Maria and Fehervary, Heleen and Collazo, Guillermo Fern{\'a}ndez and Meyns, Bart and Fresiello, Libera and Famaey, Nele},
  journal={Cardiovascular engineering and technology},
  volume={16},
  number={1},
  pages={34--51},
  year={2025},
  publisher={Springer}
}

@article{peng2018study,
  title={Study on thermally activated coiled linear actuators made from polymer fibers},
  author={Peng, Zehua and others},
  year={2018},
  publisher={Hong Kong Polytechnic University}
}

@article{aziz2023plant,
  title={Plant-like tropisms in artificial muscles},
  author={Aziz, Shazed and Zhang, Xi and Naficy, Sina and Salahuddin, Bidita and Jager, Edwin WH and Zhu, Zhonghua},
  journal={Advanced Materials},
  volume={35},
  number={51},
  pages={2212046},
  year={2023},
  publisher={Wiley Online Library}
}

@article{christ2020hydraulic,
  title={A hydraulic model of cardiovascular physiology and pathophysiology embedded into a computer-based teaching system for student training in laboratory courses},
  author={Christ, Andreas and Barowsky, Dieter and Gekle, Michael and Thews, Oliver},
  journal={Advances in Physiology Education},
  year={2020},
  publisher={American Physiological Society Bethesda, MD}
}

@article{almubarak2017twisted,
  title={Twisted and coiled polymer (TCP) muscles embedded in silicone elastomer for use in soft robot},
  author={Almubarak, Yara and Tadesse, Yonas},
  journal={International Journal of Intelligent Robotics and Applications},
  volume={1},
  pages={352--368},
  year={2017},
  publisher={Springer}
}

@article{haines2014artificial,
  title={Artificial muscles from fishing line and sewing thread},
  author={Haines, Carter S and Lima, M{\'a}rcio D and Li, Na and Spinks, Geoffrey M and Foroughi, Javad and Madden, John DW and Kim, Shi Hyeong and Fang, Shaoli and Jung de Andrade, M{\^o}nica and G{\"o}ktepe, Fatma and others},
  journal={science},
  volume={343},
  number={6173},
  pages={868--872},
  year={2014},
  publisher={American Association for the Advancement of Science}
}

@article{wang2023insect,
  title={Insect-scale jumping robots enabled by a dynamic buckling cascade},
  author={Wang, Yuzhe and Wang, Qiong and Liu, Mingchao and Qin, Yimeng and Cheng, Liuyang and Bolmin, Ophelia and Alleyne, Marianne and Wissa, Aimy and Baughman, Ray H and Vella, Dominic and others},
  journal={Proceedings of the National Academy of Sciences},
  volume={120},
  number={5},
  pages={e2210651120},
  year={2023},
  publisher={National Acad Sciences}
}

@article{ren2022stepwise,
  title={Stepwise artificial yarn muscles with energy-free catch states driven by aluminum-ion insertion},
  author={Ren, Ming and Xu, Panpan and Zhou, Yurong and Wang, Yulian and Dong, Lizhong and Zhou, Tao and Chang, Jinke and He, Jianfeng and Wei, Xulin and Wu, Yulong and others},
  journal={ACS nano},
  volume={16},
  number={10},
  pages={15850--15861},
  year={2022},
  publisher={ACS Publications}
}

@phdthesis{rafie2018characterizing,
  title={Characterizing the behavior of nylon actuators and exploring methods for manufacturing them to get the highest amount of output},
  author={Rafie Ravandi, Ali},
  year={2018},
  school={University of British Columbia}
}

@article{zhang2019modeling,
  title={Modeling and simulation of complex dynamic musculoskeletal architectures},
  author={Zhang, Xiaotian and Chan, Fan Kiat and Parthasarathy, Tejaswin and Gazzola, Mattia},
  journal={Nature communications},
  volume={10},
  number={1},
  pages={4825},
  year={2019},
  publisher={Nature Publishing Group UK London}
}

@article{tsai2023miniature,
  title={Miniature soft jumping robots made by additive manufacturing},
  author={Tsai, Samuel and Wang, Qiong and Wang, Yuzhe and King, William P and Tawfick, Sameh},
  journal={Smart Materials and Structures},
  volume={32},
  number={10},
  pages={105022},
  year={2023},
  publisher={IOP Publishing}
}

@article{van2011birth,
  title={Birth prevalence of congenital heart disease worldwide: a systematic review and meta-analysis},
  author={Van Der Linde, Denise and Konings, Elisabeth EM and Slager, Maarten A and Witsenburg, Maarten and Helbing, Willem A and Takkenberg, Johanna JM and Roos-Hesselink, Jolien W},
  journal={Journal of the American College of Cardiology},
  volume={58},
  number={21},
  pages={2241--2247},
  year={2011},
  publisher={American College of Cardiology Foundation Washington, DC}
}

@article{hoffman2002incidence,
  author       = {Hoffman, Julien I. and Kaplan, Samuel},
  title        = {The incidence of congenital heart disease},
  journal      = {Journal of the American College of Cardiology},
  year         = {2002},
  volume       = {39},
  number       = {12},
  pages        = {1890--1900},
  doi          = {10.1016/S0735-1097(02)01886-7}
}

@article{wren2012epidemiology,
  author       = {Wren, Christopher},
  title        = {The Epidemiology of Cardiovascular Malformations},
  journal      = {Pediatric Cardiology Medicine},
  year         = {2012},
  volume       = {19},
  pages        = {268},
}

@article{marelli2014lifetime,
  title={Lifetime prevalence of congenital heart disease in the general population from 2000 to 2010},
  author={Marelli, Ariane J and Ionescu-Ittu, Raluca and Mackie, Andrew S and Guo, Liming and Dendukuri, Nandini and Kaouache, Mohammed},
  journal={Circulation},
  volume={130},
  number={9},
  pages={749--756},
  year={2014},
  publisher={Lippincott Williams \& Wilkins Hagerstown, MD}
}

@article{agopian2017genome,
  title={Genome-wide association studies and meta-analyses for congenital heart defects},
  author={Agopian, AJ and Goldmuntz, Elizabeth and Hakonarson, Hakon and Sewda, Anshuman and Taylor, Deanne and Mitchell, Laura E},
  journal={Circulation: Cardiovascular Genetics},
  volume={10},
  number={3},
  pages={e001449},
  year={2017},
  publisher={Lippincott Williams \& Wilkins Hagerstown, MD}
}

@article{davies2024soft,
  title={Soft robotic artificial left ventricle simulator capable of reproducing myocardial biomechanics},
  author={Davies, James and Thai, Mai Thanh and Sharma, Bibhu and Hoang, Trung Thien and Nguyen, Chi Cong and Phan, Phuoc Thien and Vuong, Thao Nhu Anne Marie and Ji, Adrienne and Zhu, Kefan and Nicotra, Emanuele and others},
  journal={Science Robotics},
  volume={9},
  number={94},
  pages={eado4553},
  year={2024},
  publisher={American Association for the Advancement of Science}
}

@article{roche2017soft,
  title={Soft robotic sleeve supports heart function},
  author={Roche, Ellen T and Horvath, Markus A and Wamala, Isaac and Alazmani, Ali and Song, Sang-Eun and Whyte, William and Machaidze, Zurab and Payne, Christopher J and Weaver, James C and Fishbein, Gregory and others},
  journal={Science translational medicine},
  volume={9},
  number={373},
  pages={eaaf3925},
  year={2017},
  publisher={American Association for the Advancement of Science}
}

@article{weymann2023artificial,
  title={Artificial muscles and soft robotic devices for treatment of end-stage heart failure},
  author={Weymann, Alexander and Foroughi, Javad and Vardanyan, Robert and Punjabi, Prakash P and Schmack, Bastian and Aloko, Sinmisola and Spinks, Geoffrey M and Wang, Chun H and Arjomandi Rad, Arian and Ruhparwar, Arjang},
  journal={Advanced Materials},
  volume={35},
  number={19},
  pages={2207390},
  year={2023},
  publisher={Wiley Online Library}
}

@article{park2024biorobotic,
  title={Biorobotic hybrid heart as a benchtop cardiac mitral valve simulator},
  author={Park, Clara and Singh, Manisha and Saeed, Mossab Y and Nguyen, Christopher T and Roche, Ellen T},
  journal={Device},
  volume={2},
  number={1},
  year={2024},
  publisher={Elsevier}
}

@article{tsai2026hierarchical,
  title={Hierarchical Artificial Muscle with Nonlinear Elasticity for Antagonistic and Cyclic Robotics},
  author={Tsai, Samuel and Cheng, Liuyang and Albazroun, Ali and Wang, Qiong and Kim, Jeongmin and Tekinalp, Arman and Kim, Soonwook and Simcox, Charlie and Downing, Ryne and Sivaramakrishnan, Vagish and others},
  journal={Advanced Science},
  pages={e21604},
  year={2026},
  publisher={Wiley Online Library}
}

@book{formaggia2010cardiovascular,
  title={Cardiovascular Mathematics: Modeling and simulation of the circulatory system},
  author={Formaggia, Luca and Quarteroni, Alfio and Veneziani, Allesandro},
  volume={1},
  year={2010},
  publisher={Springer Science \& Business Media}
}

@article{stergiopulos1999total,
  title={Total arterial inertance as the fourth element of the windkessel model},
  author={Stergiopulos, Nikos and Westerhof, Berend E and Westerhof, Nico},
  journal={American Journal of Physiology-Heart and Circulatory Physiology},
  volume={276},
  number={1},
  pages={H81--H88},
  year={1999},
  publisher={American Physiological Society Bethesda, MD}
}

@article{genet2014distribution,
  title={Distribution of normal human left ventricular myofiber stress at end diastole and end systole: a target for in silico design of heart failure treatments},
  author={Genet, Martin and Lee, Lik Chuan and Nguyen, Rebecca and Haraldsson, Henrik and Acevedo-Bolton, Gabriel and Zhang, Zhihong and Ge, Liang and Ordovas, Karen and Kozerke, Sebastian and Guccione, Julius M},
  journal={Journal of applied physiology},
  volume={117},
  number={2},
  pages={142--152},
  year={2014},
  publisher={American Physiological Society Bethesda, MD}
}

@article{courtois1988transmitral,
  title={Transmitral pressure-flow velocity relation. Importance of regional pressure gradients in the left ventricle during diastole.},
  author={Courtois, Michael and Kov{\'a}cs Jr, Sandor J and Ludbrook, Philip A},
  journal={Circulation},
  volume={78},
  number={3},
  pages={661--671},
  year={1988}
}

@article{marsden2014optimization,
  title={Optimization in cardiovascular modeling},
  author={Marsden, Alison L},
  journal={Annual review of fluid mechanics},
  volume={46},
  number={1},
  pages={519--546},
  year={2014},
  publisher={Annual Reviews}
}

@article{bucelli2023mathematical,
  title={A mathematical model that integrates cardiac electrophysiology, mechanics, and fluid dynamics: Application to the human left heart},
  author={Bucelli, Michele and Zingaro, Alberto and Africa, Pasquale Claudio and Fumagalli, Ivan and Dede', Luca and Quarteroni, Alfio},
  journal={International journal for numerical methods in biomedical engineering},
  volume={39},
  number={3},
  pages={e3678},
  year={2023},
  publisher={Wiley Online Library}
}

@article{bonini2026monolithic,
  title={A monolithic patient-specific 3D--0D model for In silico investigation of hemodynamics in patients with left ventricular assist devices},
  author={Bonini, Mia and Hirschvogel, Marc and Ferguson, Michael and Pagani, Francis and Tang, Paul C and Nordsletten, David},
  journal={Biomechanics and Modeling in Mechanobiology},
  volume={25},
  number={3},
  pages={51},
  year={2026},
  publisher={Springer}
}

@article{guillen20263d,
  title={3D-Printed Dynamic Heart Model With Left-Side Anatomy and Integrated Sensor for Edge-to-Edge Repair and Regurgitation Reduction},
  author={Guillen Obando, Alejandro and Shen, Hongyi and McGovern, Myles and Zhang, Yusen and Lin, Vivien and Fu, Darryl and Baumwart, Ryan and Qiu, Kaiyan},
  journal={Advanced Materials Technologies},
  pages={e70885},
  year={2026},
  publisher={Wiley Online Library}
}

@article{zrinscak2025design,
  title={Design of a soft robotic artificial cardiac wall},
  author={Zrinscak, Debora and De Chirico, Claudia M and Lorenzon, Lucrezia and Coluccia, Fabiola and De Luca, Mauro and Maselli, Martina and Kluin, Jolanda and Overvelde, Johannes TB and Cianchetti, Matteo},
  journal={Artificial Organs},
  volume={49},
  number={8},
  pages={1265--1276},
  year={2025},
  publisher={Wiley Online Library}
}

@article{foroughi2026soft,
  title={Soft robotic cardiac sleeves: materials, actuation mechanisms and translational pathways},
  author={Foroughi, Javad and Aloko, Sinmisola and Spinks, Geoffrey and Wu, Liao and Wu, Shuying and Hayward, Christopher and Wang, Chun H and Ruhparwar, Arjang},
  journal={Materials Horizons},
  volume={13},
  number={11},
  pages={5237--5267},
  year={2026},
  publisher={Royal Society of Chemistry}
}

@article{ji2026soft,
  title={Soft robotic devices for cardiovascular medicine},
  author={Ji, Xing-Yu and Zhu, Jia-Qi and Wu, Ke and Wang, Liu and Xiong, Tian-Yuan and Zhang, Li and Chen, Mao},
  journal={Nature Reviews Cardiology},
  pages={1--19},
  year={2026},
  publisher={Nature Publishing Group UK London}
}

@article{kim2010patient,
  title={Patient-specific modeling of blood flow and pressure in human coronary arteries},
  author={Kim, Hyun Jin and Vignon-Clementel, IE and Coogan, JS and Figueroa, CA and Jansen, KE and Taylor, e CA},
  journal={Annals of biomedical engineering},
  volume={38},
  number={10},
  pages={3195--3209},
  year={2010},
  publisher={Springer}
}

@article{taylor2009patient,
  title={Patient-specific modeling of cardiovascular mechanics},
  author={Taylor, Charles A and Figueroa, CA},
  journal={Annual review of biomedical engineering},
  volume={11},
  number={1},
  pages={109--134},
  year={2009},
  publisher={Annual Reviews}
}

@article{nordsletten2011coupling,
  title={Coupling multi-physics models to cardiac mechanics},
  author={Nordsletten, DA and Niederer, SA and Nash, MP and Hunter, PJ and Smith, NP},
  journal={Progress in biophysics and molecular biology},
  volume={104},
  number={1-3},
  pages={77--88},
  year={2011},
  publisher={Elsevier}
}

@article{nash2000computational,
  title={Computational mechanics of the heart},
  author={Nash, Martyn P and Hunter, Peter J},
  journal={Journal of elasticity and the physical science of solids},
  volume={61},
  number={1},
  pages={113--141},
  year={2000},
  publisher={Springer}
}

@article{bazilevs2009computational,
  title={Computational fluid--structure interaction: methods and application to a total cavopulmonary connection},
  author={Bazilevs, Yuri and Hsu, M-C and Benson, David J and Sankaran, Sethu and Marsden, Alison L},
  journal={Computational Mechanics},
  volume={45},
  number={1},
  pages={77--89},
  year={2009},
  publisher={Springer}
}

@article{segers2003systemic,
  title={Systemic and pulmonary hemodynamics assessed with a lumped-parameter heart-arterial interaction model},
  author={Segers, Patrick and Stergiopulos, Nikos and Westerhof, Nico and Wouters, Patrick and Kolh, Philippe and Verdonck, Pascal},
  journal={Journal of engineering mathematics},
  volume={47},
  number={3},
  pages={185--199},
  year={2003},
  publisher={Springer}
}

@article{kim2009coupling,
  title={On coupling a lumped parameter heart model and a three-dimensional finite element aorta model},
  author={Kim, Hyun Jin and Vignon-Clementel, Irene E and Figueroa, C Alberto and LaDisa, John F and Jansen, Kenneth E and Feinstein, Jeffrey A and Taylor, Charles A},
  journal={Annals of biomedical engineering},
  volume={37},
  number={11},
  pages={2153--2169},
  year={2009},
  publisher={Springer}
}

@article{shi2011review,
  title={Review of zero-D and 1-D models of blood flow in the cardiovascular system},
  author={Shi, Yubing and Lawford, Patricia and Hose, Rodney},
  journal={Biomedical engineering online},
  volume={10},
  number={1},
  pages={33},
  year={2011},
  publisher={Springer}
}

@article{polygerinos2015modeling,
  title={Modeling of soft fiber-reinforced bending actuators},
  author={Polygerinos, Panagiotis and Wang, Zheng and Overvelde, Johannes TB and Galloway, Kevin C and Wood, Robert J and Bertoldi, Katia and Walsh, Conor J},
  journal={IEEE Transactions on Robotics},
  volume={31},
  number={3},
  pages={778--789},
  year={2015},
  publisher={IEEE}
}

@article{timms2005complete,
  title={A complete mock circulation loop for the evaluation of left, right, and biventricular assist devices},
  author={Timms, Daniel and Hayne, Mark and McNeil, Keith and Galbraith, Andrew},
  journal={Artificial organs},
  volume={29},
  number={7},
  pages={564--572},
  year={2005},
  publisher={Wiley Online Library}
}

\newpage
\begin{appendices}

\renewcommand{\thefigure}{S\arabic{figure}}

\setcounter{figure}{0}

\section{Twisted-and-Coiled Polymer Actuator (TCPA) Fabrication}\label{Appendix1}
Nylon monofilaments (1.0 mm and 1.3 mm diameters) were purchased from Shaddock Fishing, and 0.15 mm nylon fibers were obtained from Beadalon. Electrical wires for Joule heating were supplied by Consolidated Electronic Wire and Cable.

The TCPA fabrication protocol followed the methodology established in our previous work \cite{tsai2026hierarchical}. Fabrication of monocoiled muscles (M-1) was initiated by twisting 1.0 mm fibers under a tensile load of 800 g, followed by coiling around mandrels of various diameters (Table~\ref{Table1}). For the 1.3 mm fiber muscles (M-1.3), an increased load of 1300 g was applied during twisting. Supercoiled muscles were manufactured by twisting three nylon fibers and an electrical wire together using a three-channel connector. The resulting 3-ply structure was then mandrel-coiled. For S-0.47 muscles, twisting weights of 600 g (nylon) and 200 g (electrical wire) were applied, with a coiling weight of 1500 g. For S-0.15 muscles, these weights were adjusted to 60 g, 100 g, and 140 g, respectively. Hypercoiled muscles (H-0.47-0.32) utilized 3-ply S-0.47 structures; these were annealed at 170°C for 2 hours, twisted again under 1500 g, and subsequently mandrel-coiled.

All muscles, with the exception of S-0.15-0.25, were annealed at 170°C for 2 hours in peanut oil, stretched to introduce inter-coil gaps, and annealed for an additional 6 hours. The S-0.15-0.25 fibers were annealed at 170°C for 1 hour, stretched, and annealed for an additional hour. 

\section{Extended Analysis of the Isometric Test}\label{Appendix2}

\begin{figure}[tp!]
    \centering
    \includegraphics[width=1\textwidth]{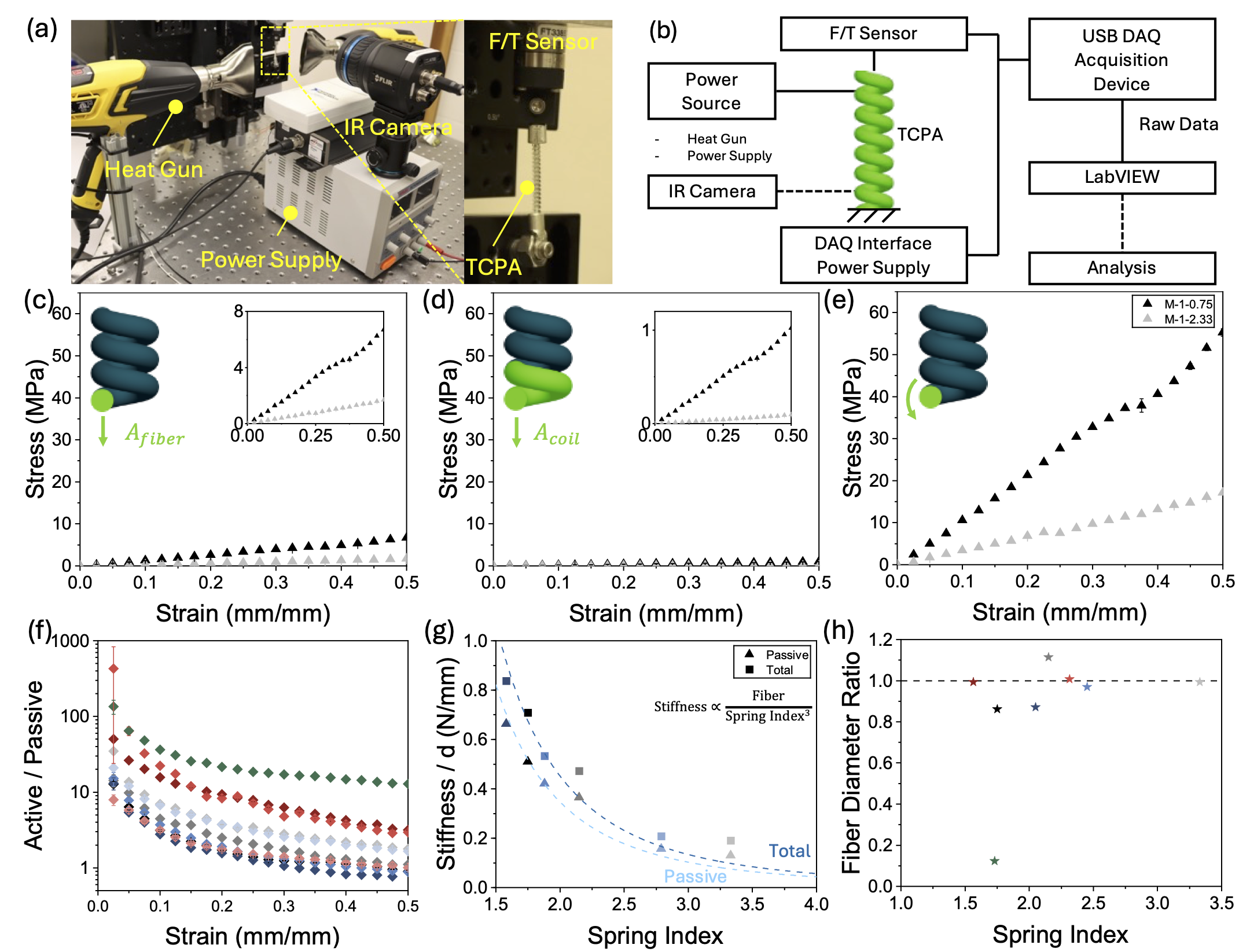}
    \caption{\textbf{Details of the isometric test and additional analysis.} (a) Hardware system of the isometric test. (b) Flowchart to process the data. (c) Normal stress based on fiber area. (d) Normal stress based on coil area. (e) Maximum shear stress. (f) Ratio between active force and passive force. (g) Slope of passive and active force versus spring index of TCPAs. (h) Ratio between the estimated fiber diameters of equivalent helical spring based on passive force and the muscle diameters.}
    \label{fig:FigureS1}
\end{figure}

Fig.~\ref{fig:FigureS1}a--b illustrate the experimental setup, along with the data acquisition and assessment procedures. Throughout the tests, the surface temperature was monitored using an infrared (IR) camera. Depending on the specific muscle type, thermal actuation was provided by either a pair of heat guns or an electrical power supply.

Different approaches for calculating the internal stress of TCPAs were also evaluated. While extensive research has focused on the stroke and work capacity of TCPAs, a clear consensus on defining their internal stress is still lacking. Fig.~\ref{fig:FigureS1}c--e illustrates monocoiled muscles fabricated from nylon fibers of identical diameter but coiled around mandrels of varying sizes, evaluated using three distinct cross-sectional stress definitions: normal stress based on the fiber area, normal stress based on the coil area, and maximum shear stress. The corresponding equations are defined as:

\begin{equation}
    \sigma_a = \frac{F}{A_\text{fiber}} = \frac{4F}{\pi d^2}
    \label{Equation_B1}
\end{equation}

\begin{equation}
    \sigma_b = \frac{F}{A_\text{coil}} = \frac{4F}{\pi D^2}
    \label{Equation_B2}
\end{equation}

\begin{equation}
    \tau_c = \frac{8Fd}{\pi D^3}\left(\frac{4S-1}{4S-4}+\frac{0.615}{S}\right)
    \label{Equation_B3}
\end{equation}

where $d$ is the fiber diameter, $D$ is the mean coil diameter, and $S$ represents the spring index ($S = D/d$). As depicted, Fig.~\ref{fig:FigureS1}d represents the lower bound of the stress estimation, whereas Fig.~\ref{fig:FigureS1}e provides the upper bound. Calculating normal stress based on the fiber cross-sectional area offers a direct measure of intrinsic strength, but it fails to account for the influence of the mandrel diameter on the spring index. Conversely, defining normal stress using the coil cross-sectional area incorporates macroscopic geometric effects but significantly underestimates the actual stress by neglecting shear and torsional components. In contrast, the maximum shear stress approach yields a more realistic estimation of the stresses experienced by the TCPAs, accurately reflecting the non-uniform stress distribution. Therefore, the incorporation of shear stress is essential for a comprehensive understanding of the mechanical limits of TCPAs.

Furthermore, Fig.~\ref{fig:FigureS1}f presents the active-to-passive force ratio, which indicates the capacity of the muscle to overcome passive resistance within a coupled system. Notably, although the H-0.47-0.32 actuator generated the lowest absolute force, it exhibited the highest active-to-passive force ratio.

Additionally, the mono-fiber TCPAs were modeled as conventional helical springs to calculate their equivalent diameters, as expressed in the following equations:

\begin{equation}
    k_\text{exp} = \frac{\sigma A}{\varepsilon L} = \frac{F}{\varepsilon L} = \frac{E^*}{L} \label{Equation_B4}
\end{equation}

\begin{equation}
    k_\text{est} = \frac{G d_\text{sim}^4}{8 N (d_\text{sim} + m)^3}
    \label{Equation_B5}
\end{equation}

where $E^*$ denotes the slope of each curve from Fig.~\ref{fig:Figure5}d, while $A$ and $L$ represent the cross-sectional area and length of the muscles, respectively. The variable $N$ indicates the number of coils, measured directly from the fabricated TCPAs. By equating $k_\text{exp}$ and $k_\text{est}$, the equivalent simulated diameter, $d_\text{sim}$, was determined.

Based on this mathematical estimation, Fig.~\ref{fig:FigureS1}g illustrates the fitted relationship between stiffness and the spring index. The analysis revealed that coiled muscles with a larger spring index generated lower passive and active forces at an equivalent strain. Monocoiled muscles and the S-0.15-0.25 variant consistently exhibited higher passive forces relative to their active forces, indicating a dominance of structural resistance. Conversely, the active forces of the S-0.47 and H-0.47-0.32 muscles exceeded their corresponding passive counterparts. Furthermore, the force-strain slopes for the monocoiled muscles were inversely proportional to the cube of the spring index, which aligns well with classical engineering theories for spring stiffness.

The obtained passive force values were utilized to estimate the equivalent fiber diameters of the monocoiled TCPAs. As shown in Fig.~\ref{fig:FigureS1}h, the simulated fiber diameters demonstrated strong agreement with the actual measured diameters for the majority of the monocoiled TCPAs. However, the hypercoiled muscle exhibited an exceptionally low ratio of actual-to-estimated diameter. This discrepancy is likely attributable to its larger cross-sectional area and complex inter-ply frictional interactions, which remain challenging to accurately capture through analytical estimation.

\section{Numerical Analysis and Comparison of Isobaric and Isometric Tests}\label{Appendix3}
\begin{figure}[tp!]
    \centering
    \includegraphics[width=1\textwidth]{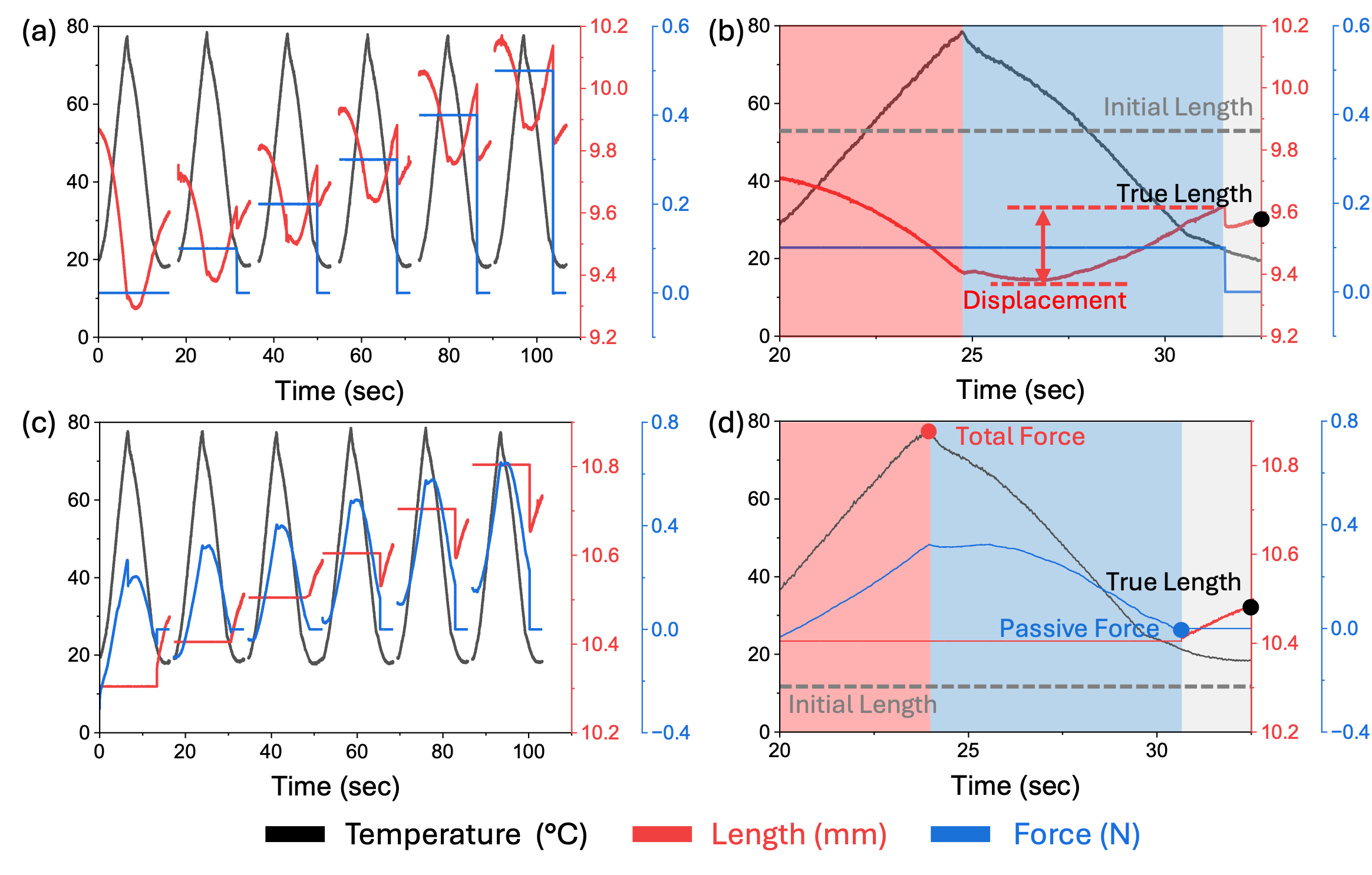}
    \caption{\textbf{Isobaric and isometric tests on dynamic mechanical analysis (DMA).} (a) Full process of the isobaric test. (b) A representative single cycle of the isobaric test. (c) Full process of the isometric test. (d) A representative single cycle of the isometric test.}
    \label{fig:FigureS2}
\end{figure}

The first cycle of each test was excluded from the analysis as an outlier due to inconsistent results caused by initial viscoelastic deformation. In both tests, the initial length was taken from the very first measurement. The final length of each cycle was considered as the true length, which was the neutral length after exposure to high temperature under given circumstances: strain or force. For the second cycle of the isobaric test (Fig. \ref{fig:FigureS2}a), the displacement was the difference between the minimum length and the length immediately after reaching the minimum temperature (Fig. \ref{fig:FigureS2}b). In the isometric test (\ref{fig:FigureS2}c), the passive force was recorded as the last point of the cooling step, while the total force was defined as the maximum force within the cycle (Fig. \ref{fig:FigureS2}d).

We compared the results of the isometric test with those of the isobaric test using an identical S-0.47-0.75 muscle with the same length and energy input. When using F/T sensor and displacement sensor, the energy input was regulated by controlling the applied current and duration. To minimize the impact of creep and stress relaxation during the testing, the maximum temperature was restricted to 80 °C. We calculated the adjusted neutral length after each measurement in both the isobaric and isometric tests in order to consider the effect of deformations. In the isometric test, we adjusted the stretched length of the muscles so that the weight of muscles was included in the measured forces in between two displacements. We then introduced interpolation to calculate the force with respect to the displacement and obtained the stretched equilibrium length. This procedure allowed us to calculate the normal and adjusted strain or stroke by dividing displacement by initial neutral length and adjusted neutral length, respectively. The strain and stroke share the same equation for the definition, which is the following.

\begin{equation}
    \text{Strain (Stroke)} = \frac{\text{Final Length - Initial Length}}{\text{Initial Length}}
    \label{Equation_C6}
\end{equation}

While the strain in the isometric test has a positive sign as it is in the tensional region, the stroke is defined as negative due to contractile motion implying a shorter final length with respect to the initial length.

The test was repeated on the dynamic mechanical analyzer (DMA 850; TA Instruments) to justify the results from our new isometric testing setup. The maximum temperature was limited to 80 °C for the same reason. The detailed information is included in the supplementary material. The tensile tests were conducted before and after each test to see if there was a significant change in mechanical properties of TCPAs.

\begin{figure}[tp!]
    \centering
    \includegraphics[width=1\textwidth]{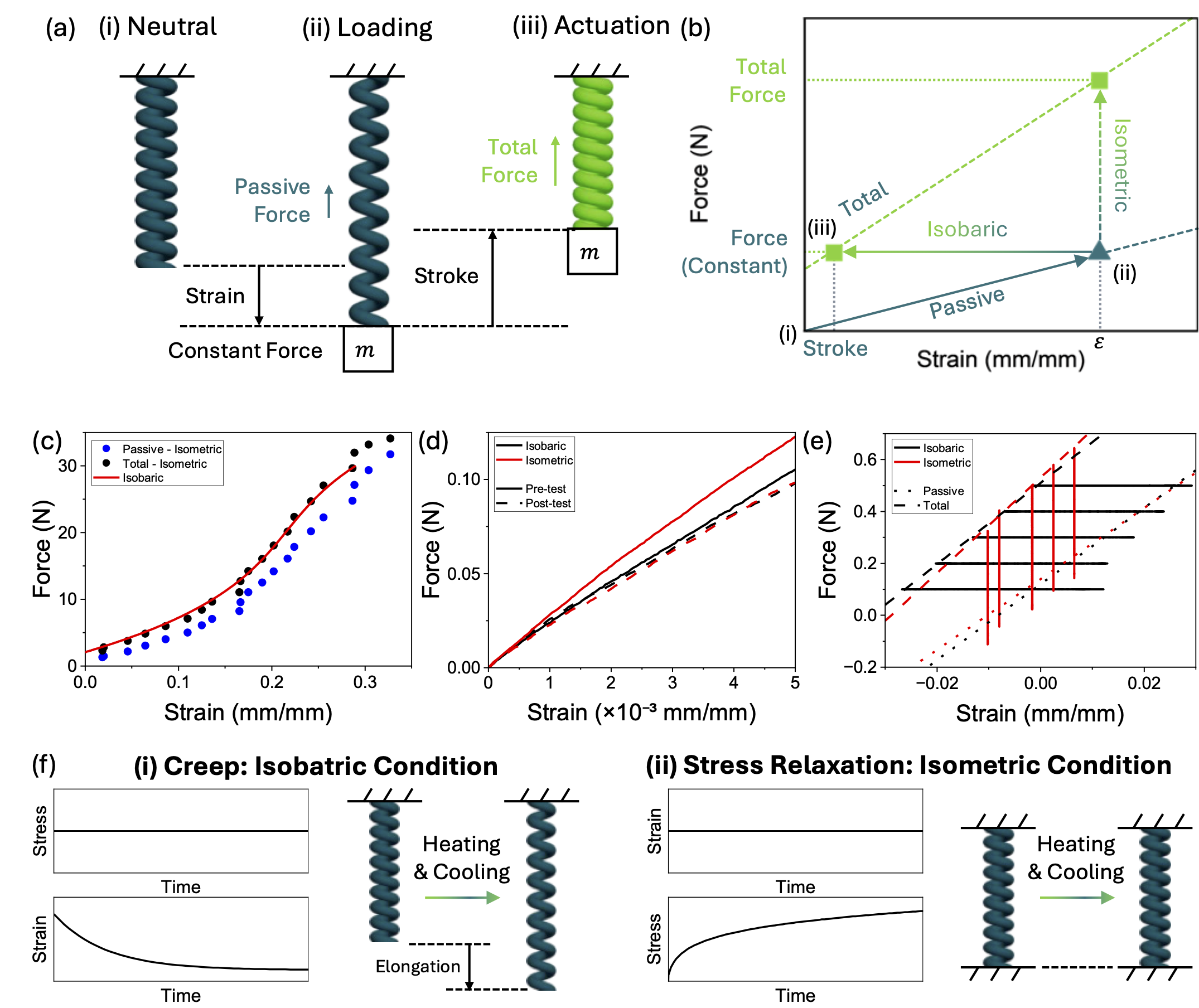}
    \caption{\textbf{Equivalence of the isometric and isobaric tests.} (a) Scheme of the conventional isobaric test. (b) Flow difference of the isobaric test from the isometric test in the force-strain plot. (c) The approximated isobaric curve and the experimental isometric curves of S-0.47-0.75. (d) Tensile test of S-0.47-0.75 before and after the isobaric and isometric tests. (e) Comparison between the isobaric and isometric curves of S-0.47-0.75. (f) Scheme of creep and stress relaxation and their effects on the length of the TCPAs.}
    \label{fig:FigureS3}
\end{figure}

Fig.~\ref{fig:FigureS3}a--b show the difference between the conventional isobaric test and the isometric test. In (ii) the loading step, a known force is applied to the TCPA, causing it to stretch (Fig.~\ref{fig:FigureS3}a). Since it is not constrained in length, it contracts during actuation by transferring from the passive curve to the total curve horizontally, whereas the actuation step is represented vertically in the isometric test (Fig.~\ref{fig:FigureS3}b).

In order to evaluate the correlation between the isobaric test and the isometric test, we aimed to derive the total force as a function of strain from the isobaric test. For this purpose, we used the passive force obtained from the isometric test to calculate the predicted active force. By adding the calculated active force and the passive force, we obtained the fitted total force from the isobaric test for both the normal strain and adjusted strain. As demonstrated in Fig.~\ref{fig:FigureS3}c, the fitted values from the isobaric test were well consistent with the total force from the isometric test. These results suggest that the isobaric and isometric tests are fundamentally equivalent in their ability to measure force-strain characteristics. Moreover, the isometric test proves to be helpful in identifying the most suitable muscles for specific applications that contain elastic components, because it can differentiate the active term from the passive force.

We then investigated the intrinsic equivalence between the isometric and isobaric tests, performed on the same instrument under the same conditions, by limiting the test region to extremely small strains. We took data points where muscles were fully actuated at the maximum target temperature of each test (Fig.~\ref{fig:FigureS3}d--e). Fig.~\ref{fig:FigureS3}d shows the tensile test before and after each test to validate if the muscle experienced a huge amount of nonrecoverable stress relaxation or creep. For the isobaric test, the slope changed from 17.18 N/(mm/mm) to 16.60 N/(mm/mm). In the isometric test, it also decreased from 19.12 N/(mm/mm) to 18.79 N/(mm/mm). We extracted the length so that we can calculate the true strain or stroke based on the deformed neutral length by creep and soft relaxation. Most importantly, we took the passive and total stiffness as from each test. The passive stiffness of the isobaric test was 13.66 N / (mm / mm), while that of the isometric test was 15.66 N/(mm/mm). Also, the isobaric and isometric tests yielded 14.65 N/(mm/mm) and 18.38 N/(mm/mm) of total stiffness, respectively. This indicates that the muscle softened to a small extent, but it was negligible as the slope changes were few. Furthermore, since the slopes taken from each test were extremely similar to each other, we can expect the muscle to undergo a similar degree of changes in mechanical properties in both methods.

We would like to note the difference in the mechanisms between the isometric and isobaric actuation; the active modulus was positive in the isometric test but dynamically shifted in the isobaric test. One of the fundamental reasons is based on the material property that nylon is a viscoelastic material. Specifically, it experiences stress relaxation under the isometric condition and creep under the isobaric condition (Fig.~\ref{fig:FigureS3}f). In the isometric test, as the muscle length is fixed, polymer chain mobility plays a more significant role. Polymer chains rearrange under high temperatures to reduce internal stress, leading to an increase in active force with higher strain by enabling the muscle to store more elastic energy at the given strain \cite{peters2018effect, wang2024mechanics}. On the other hand, as the muscle length is free in the isobaric test, elongation is the dominant factor \cite{tsai2025high}, which results in less working performance. 

In addition, the differences in physical outcomes observed between the isometric and isobaric tests are rooted in the fundamental working principles of TCPAs investigated across macro to nanoscale \cite{wang2024mechanics, sharafi2015multiscale, yang2016top}. Thermal actuation involves multiple stress components: elastic storage stress (microscale) \cite{sharafi2015multiscale, yang2016top}, torsional-to-axial coupling stress (mesoscale) \cite{wang2023insect, wang2024mechanics}, and radial stiffening stress (macroscale) \cite{wang2024mechanics}. In principle, thermal actuation generates internal stresses, which are released in the form of stroke under the isobaric condition. In contrast, under the isometric condition, the additional force buildup that would otherwise be released through contraction is directly measured. Therefore, the tendency of internal stresses to be converted into mechanical work results strictly in force buildup when length is constrained. In the context of a soft robotic heart, this stored mechanical energy acts as the primary driving force that smoothly bridges the IVC into the explosive ventricular ejection phase once the aortic valve opens.

\section{Numerical and experimental performance of the antagonistic structure}\label{Appendix4}

\begin{figure}[tp!]
    \centering
    \includegraphics[width=1\textwidth]{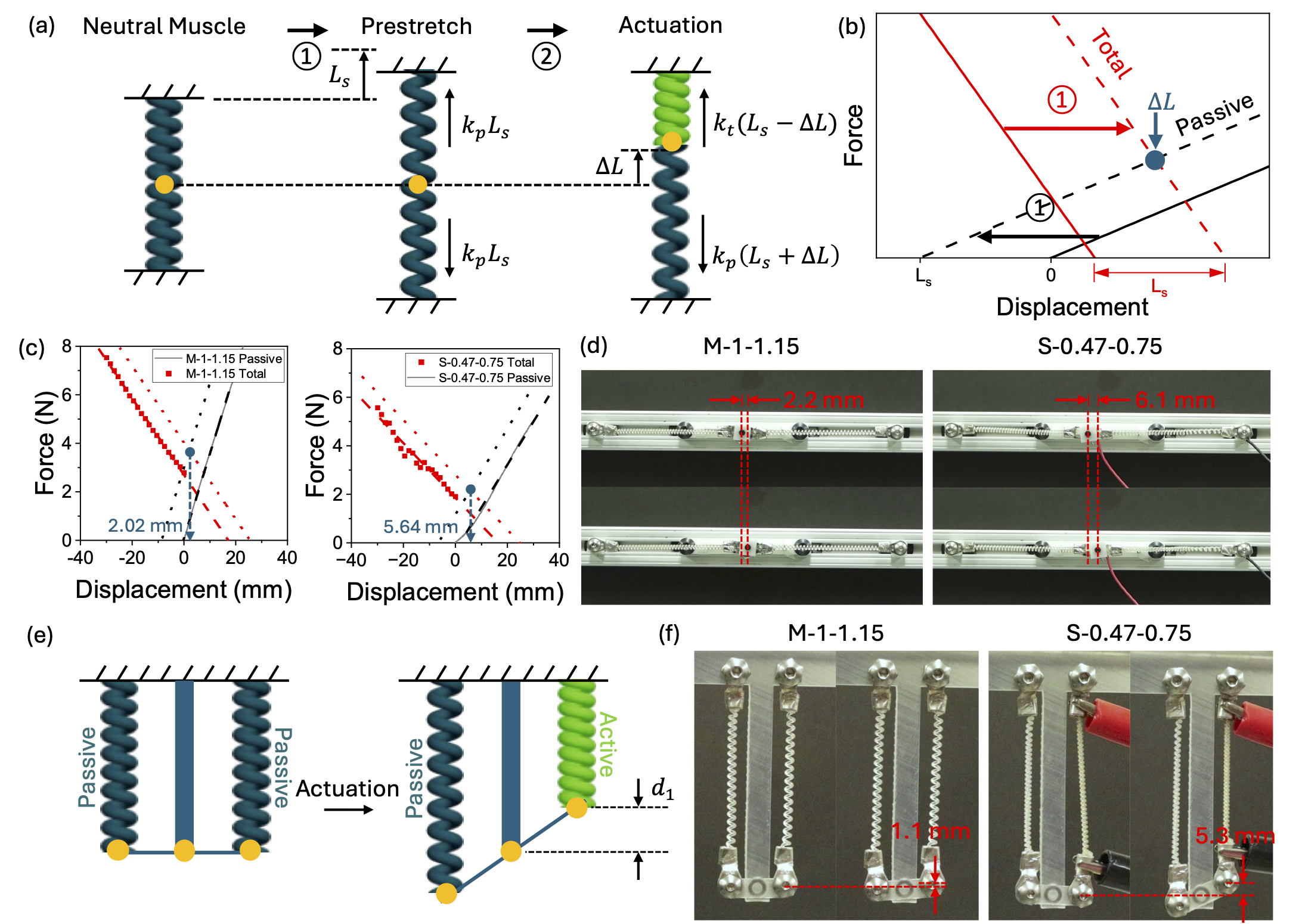}
    \caption{\textbf{TCPA performance on antagonistic structures.} (a) Scheme of translational antagonistic structure. (b) Prediction of stroke in antagonistic structures using passive total curves. (c) Prediction of M-1-1.15 and S-0.47-0.75. (d) Translational results of M-1-1.15 and S-0.47-0.75. (e) Scheme of rotational antagonistic structure. (f) Rotational results of M-1-1.15 and S-0.47-0.75.}
    \label{fig:FigureS4}
\end{figure}

The notable advantage of the isometric test is its ability to facilitate a comprehensive understanding of TCPA mechanics within interconnected structures. This predictive capability becomes particularly relevant when considering the application of TCPAs in soft robots, where multiple modules are integrated into flexible elastomeric frames to achieve complex motions \cite{tsai2023miniature}. While previous studies have explored how TCPAs contribute to various actuation modes \cite{almubarak2017twisted}, there remains a significant gap in systematically unraveling their active-passive mechanical interactions. As demonstrated in Fig.~\ref{fig:FigureS4}a--b, the isometric test enables us to analytically predict the maximum achievable stroke by explicitly mapping the actuator's total force generation against the known passive resistance of external structures.

To evaluate the performance of muscles within a system that incorporates these passive components, we conducted simulations to predict the maximum displacement achievable by the active muscle. The relative magnitude of active forces to passive forces was investigated through these simulations, where the passive forces were obtained from tensile tests (Universal Testing Machine, Instron™ ElectroPuls E1000, UK) to avoid thermal softening effects. Detailed information about the simulation procedure is included in the supplementary information. Based on the quantitative simulation results mapping the total and passive forces (Fig.~\ref{fig:FigureS4}c), it was anticipated that M-1-1.15 would achieve a maximum displacement of 2.02 mm, while S-0.47-0.75 would achieve a displacement of 5.64 mm.

To validate these predictions experimentally, translational antagonistic structures (Design 1) consisting of two identical muscles in series were designed. In this setup, one muscle was actuated while the other remained unactuated, allowing the evaluation of the effective contributions of active forces in the presence of passive forces. The frame for the antagonistic structure was fabricated by Form 3+ using Rigid 4000 resin (Formlabs Inc., USA). During thermal actuation, the connection part of the system translated following the contraction of the active muscle, resulting in a displacement denoted as $d_\text{0}$. As shown in Fig.~\ref{fig:FigureS4}d, the achieved displacement $d_\text{0}$ for M-1-1.15 was 2.2 mm, while S-0.47-0.75 exhibited a displacement of 6.1 mm. Notably, these experimental results closely aligned with the anticipated values from the simulations with an error of less than 10\%.

Furthermore, we explored a rotational geometry (Design 2) using a ball bearing, as illustrated in Fig.~\ref{fig:FigureS4}e. However, even though we used the exact same muscles, M-1-1.15 and S-0.47-0.75 only contracted 1.1 mm and 5.3 mm, respectively (Fig.~\ref{fig:FigureS4}f). The reduced performance of Design 2 compared to the translational design was attributed to the mechanical friction at the bearing (see Supplementary materials). Considering this friction, both Design 1 and Design 2 displayed principally consistent contractile behaviors. In light of these findings, the displacement achieved by the translational configuration (Design 1) serves as a more reliable indicator of intrinsic muscle performance in interconnected systems, as it is significantly less affected by external frictional losses. Consequently, the isometric testing framework proves to be a highly valuable tool for comprehending these intricate dynamics, ultimately advancing the systematic development and optimization of soft robotic systems for diverse real-world applications.

\section{Additional Information of SEA-driven Soft Robotic Ventricle}\label{Appendix5}

\begin{figure}[tp!]
    \centering
    \includegraphics[width=1\textwidth]{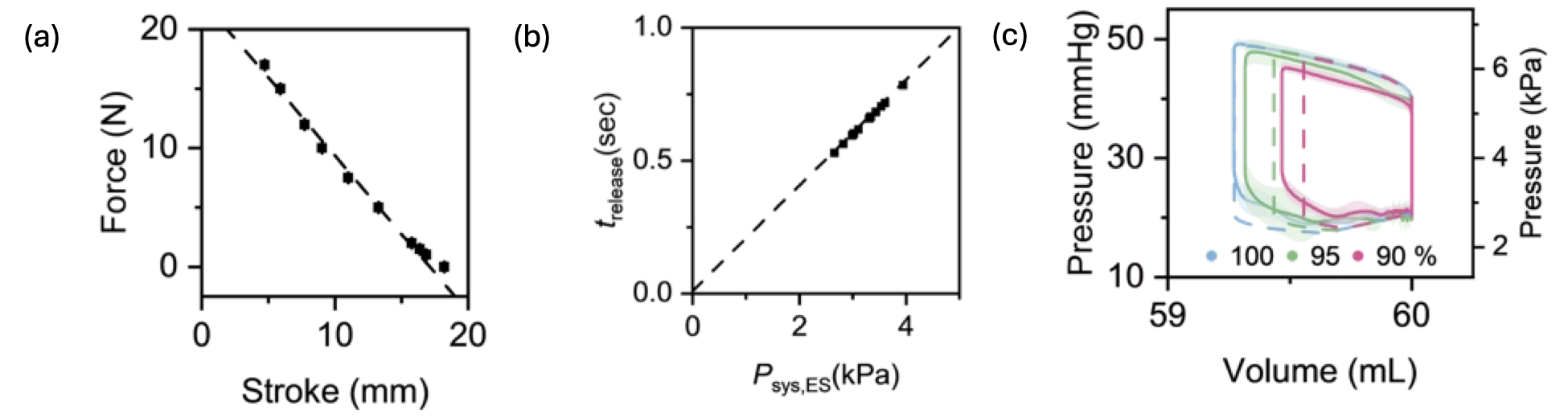}
    \caption{Motor characterization and control of actuator performance via pulse width modulation (PWM). (a) Force-stroke relationship of the linear actuator during an isobaric test with 1.5 seconds contraction duration. (b) Release time ($t_{\text{release}}$) of $F_{\text{motor}}$ as a function of ventricular pressure at end-systole ($P_{\text{sys,ES}}$). (c) Pressure-volume loops demonstrating the PWM control of the SEA-driven soft pump.}
    \label{fig:FigureS5}
\end{figure}

The performance of the linear actuator used in the SEA-driven soft robotic ventricle was evaluated by measuring the stroke under varying external loads. The capability of the stepper motor is controlled by pulse width modulation (PWM), which ranges from 0 to 255 based on the analog signal. We fixed the PWM at 255 and operated the contraction for 1.5 seconds to match the duration of systole in the soft robotic ventricle. Under these conditions, the advance of the actuator was limited not by the available force but by the short active duration, and the extrapolated zero-load value of $x_\text{act}$ reached 17.16~mm (Fig.~\ref{fig:FigureS5}a). For the pre-displacements tested ($x_\text{ED} = 15$, $17.5$ and $20$~mm), the travel remaining before the mechanical limit is $L_\text{max} - x_\text{ED} \le 15$~mm, which is shorter than this value in every case. The equality branch of Eq.~\ref{Equation_15} is therefore reached in all pre-displacement conditions, and it is this kinematic arrest that constrains the P--V loops in Fig.~\ref{fig:Figure4}b. Once the actuator is arrested, the displacement rather than the force determines the equilibrium, so the chamber compression follows the third regime of Eq.~\ref{Equation_17}, $x_\text{c} = k_\text{s}\left(L_\text{max} - x_\text{ED}\right) / \left(k_\text{s} + k_\text{c} + \varphi_\text{M} k_\text{p}\right)$, and is independent of the residual spring force stored at end-diastole; the surplus motor output is reacted by the end stop rather than transmitted to the fluid.

At the beginning of diastole the linear actuator releases quickly and then remains at a fixed displacement for the remainder of the diastolic interval. The release duration ($t_\text{release}$) scales linearly with the end-systolic ventricular pressure ($P_\text{sys,ES}$) and extrapolates through the origin (Fig.~\ref{fig:FigureS5}b), as expected physically since no actuation implies no release. The diastolic driving pressure therefore decays linearly at a fixed rate, 

\begin{equation}
    P_\text{act,dia}(t) = \max\!\left(0,\; P_\text{sys,ES} - S\,t\right)
    \label{Equation_E7}
\end{equation}

where $S$ is the reciprocal of the slope in Fig.~\ref{fig:FigureS5}b, and the release completes at
$t_\text{release} = P_\text{sys,ES}/S$.

The inotropy, or intrinsic contractility, of the soft robotic pump can be systematically modulated by adjusting the input energy of the actuators. For the SEA-driven baseline the motor force is proportional to the PWM duty cycle, so reducing the duty cycle from 100~\% (PWM~$=255$) to 90~\% (PWM~$=232$) scales $F_\text{motor}$ proportionally and shifts the force--stroke characteristic of Fig.~\ref{fig:FigureS5}a downward. In this configuration, the end-systolic displacement is set by the balance between the reduced motor force and the series stiffness. The stroke therefore decreases in proportion to the duty cycle, and with it the chamber compression $x_\text{c}$ and the active pressure $P_\text{act} = k_\text{c} x_\text{c} / A_\text{chamber}$. This produces a simultaneous reduction in peak systolic pressure and stroke volume (Fig.~\ref{fig:FigureS5}c).

\section{Additional Information of TCPAs-driven Soft Robotic Ventricle}\label{Appendix6}

Equation~\ref{Equation_24} scales the peak isometric force by the normalized thermal response $T(t)$, whose functional form is derived here. Joule heating raises the TCPA temperature during systole and forced convection removes it during diastole; because the polymer transduces temperature into force through its thermal expansion, both phases follow first-order thermal relaxation with distinct time constants. Normalizing so that $T = 1$ at end-systole and $T = 0$ at end-diastole gives

\begin{equation}
    T(t) =
    \begin{cases}
        \dfrac{1-e^{-t/\tau_\text{h}}}{1-e^{-t_\text{sys}/\tau_\text{h}}}
            & 0 \le t \le t_\text{sys} \\[10pt]
        \dfrac{e^{-\left(t-t_\text{sys}\right)/\tau_\text{c}}
              -e^{-\left(t_\text{cycle}-t_\text{sys}\right)/\tau_\text{c}}}
              {1-e^{-\left(t_\text{cycle}-t_\text{sys}\right)/\tau_\text{c}}}
            & t_\text{sys} < t \le t_\text{cycle}
    \end{cases}
    \label{Equation_F8}
\end{equation}

where $\tau_\text{h}$ and $\tau_\text{c}$ are the heating and cooling time constants and
$t_\text{dia}$ is the diastolic duration. The two branches are continuous at $t=t_\text{sys}$,
where both equal unity, and the normalization ensures that the cycle closes at $T=0$ irrespective
of the ratio $t_\text{sys}/\tau_\text{h}$ or $t_\text{dia}/\tau_\text{c}$. Fitting the measured
pressure waveforms gives $\tau_\text{h}=5.19$~s and $\tau_\text{c}=7.62$~s; cooling is the slower
process even with forced convection, which sets the minimum achievable cycle period of the
TCPA-driven configuration.

\begin{figure}[tp!]
    \centering
    \includegraphics[width=1\textwidth]{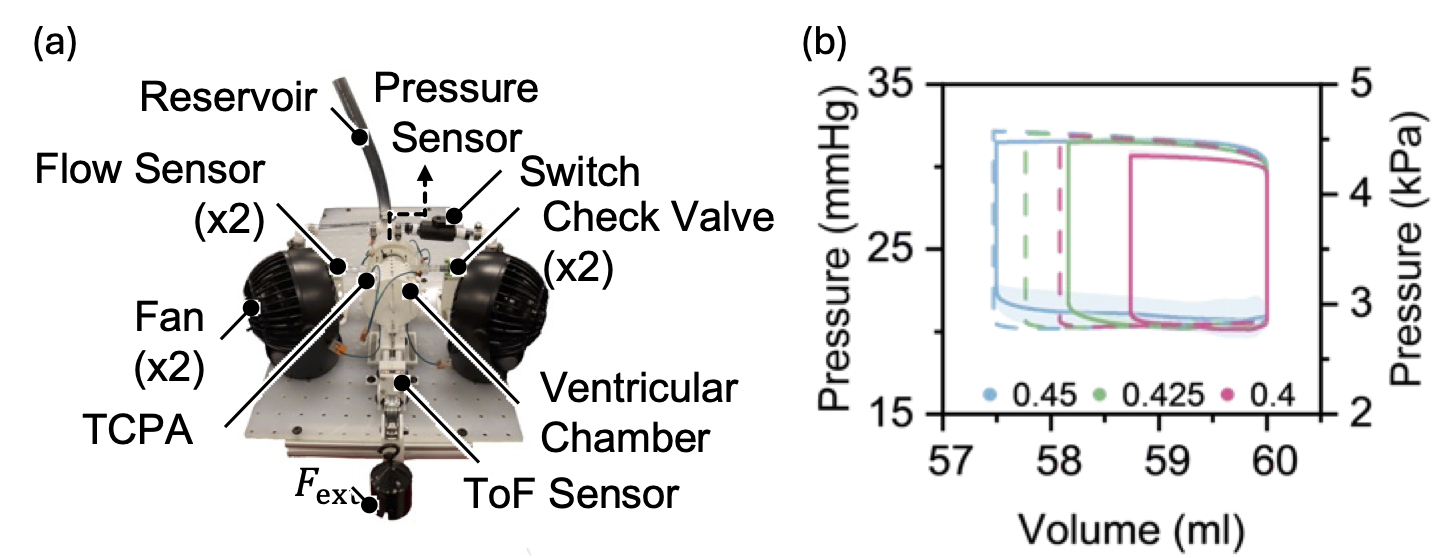}
    \caption{TCPA-driven soft robotic ventricle setup and input energy variation test. (a) Experimental setup. The ventricular chamber is actuated by $\varphi_\text{M}$ twisted-and-coiled polymer actuators (TCPAs) mounted in parallel around it, and is loaded by a dead weight providing the external diastolic load $F_\text{ext}$. Two check valves enforce unidirectional flow between the chamber and the reservoir, and two flow sensors measure the inflow and outflow independently. A pressure sensor records the intraventricular pressure, and a time-of-flight (ToF) sensor tracks the chamber displacement. Two fans provide forced convection to accelerate cooling during diastole, and a switch gates the Joule heating current supplied to the TCPAs. (b) Pressure--volume loops for three Joule heating currents, $I=0.45$, $0.425$ and $0.40$~A. Solid lines are the cycle-averaged experimental loops, and dashed lines the corresponding simulations. Lowering the heating current reduces both the peak systolic pressure and the stroke volume, providing an electrically controlled analogue of inotropic modulation.}
    \label{fig:FigureS6}
\end{figure}

The amplitude of the actuation is set by the Joule heating current $I$. The dissipated power
scales as $I^{2}R$, and for a fixed heating duration and thermal resistance the steady-state
temperature rise above ambient is proportional to it. Since both the blocked force and the
post-activation stiffness scale with the fraction of the polymer that has undergone the thermal
transition, they are modulated by a common factor

\begin{equation}
    \gamma_I = \left(\frac{I}{I_\text{ref}}\right)^{2},
    \qquad
    F_\text{M,0} \rightarrow \gamma_I F_\text{M,0},
    \qquad
    k_\text{t} \rightarrow \gamma_I k_\text{t}
    \label{Equation_F9}
\end{equation}

where $I_\text{ref}=0.45$~A is the reference current at which the actuators were characterized
($\gamma_I=1$). The passive stiffness $k_\text{p}$ is measured below the transition and is
therefore taken as temperature-independent. Substituting Eqs.~\ref{Equation_F8} and
\ref{Equation_F9} into Eq.~\ref{Equation_24} yields the driving pressure used in the
Windkessel simulation. Because $\gamma_I$ appears in both the numerator and the denominator of
$\delta_\text{ES}$, the predicted sensitivity to the heating current is weaker than measured: over
the range tested the model reproduces the direction of the change in peak pressure and stroke
volume (Fig.~\ref{fig:FigureS6}b) but underestimates its magnitude, indicating that the
thermal transition affects the active force more strongly than the post-activation stiffness.

\end{appendices}

\end{document}